\documentclass[prd,twocolumn,draft]{revtex4}
\usepackage{bm}
\usepackage{hyperref}
\usepackage{latexsym}
\usepackage{mathrsfs}
\usepackage{graphicx}
\usepackage{graphics}
\usepackage{txfonts}
\usepackage{fancybox}
\usepackage{helvet}
\usepackage{fancyhdr}
\usepackage{color}
\input epsf.sty
\usepackage{IEEEtrantools}
{\renewcommand{\theequation}{\textcolor{blue}{\arabic{equation}}}

\begin{document}

\font\titlefont=cmssbx16 
\font\subtitlefont=cmssbx9
\font\authorfont=cmr10
\font\abstractfont=cmr9
\font\abstractit=cmti9
\font\abstractbf=cmbx9
\font\sectionfont=cmssbx8
\font\subsectionfont=cmssbx8
\font\subsubsectionfont=cmssbx7
\font\figtitlefont=cmssbx7
\font\figtextfont=cmss7
\font\tabletitlefont=cmssbx7 
\font\tabletextfont=cmss7
\font\reftitlefont=cmssbx7 at 12pt
\font\reftextfont=cmr7
\font\refit=cmti7
\font\refbf=cmbx7
\font\headerfont=cmss9
\font\headerit=cmssi9
\font\footerfont=cmss9

\title{{\titlefont Lepton Generation Problem, \\Some Properties and Implications of the \\Curved Spacetime Dirac Equation.\\{\subtitlefont 
 (Curved Spacetime Dirac Equation II)}
}
}

\author{\authorfont \textbf{\textsc{Golden Gadzirayi Nyambuya}}}
\email{gadzirai@gmail.com}


\date{\today}

\begin{abstract}
{\small
\begin{center}

\end{center}
\linethickness{1.5pt}
\line(1,0){400}\\

\noindent $\textbf{\Large {\subsectionfont Abstract.}}$  This reading is a continuation of the earlier reading Nyambuya ($2008$); where three new  Curved Spacetime Dirac Equations  have been derived  mainly to try and account in a natural way for the observed anomalous gyromagnetic ratio of fermions and the suggestions is that particles including the Electron, which is thought to be a point particle, do have a finite spatial size and this is one of the reasons for the observed anomalous gyromagnetic ratio. Combining the idea in Nyambuya ($2008$) which lead to the derivation of the three new  Curved Spacetime Dirac Equations, and the proposed Unified Field Theory (Nyambuya $2007$), a total of $12$ equations each with $16$ sub-components are generated thus leading to a total of $192$ equations for the Curved Spacetime Dirac Equation. Some symmetries of these equations are investigated, \textit{i.e.}, the Lorentz symmetry, charge conjugation symmetry  ($ \textrm{C}$), time reversal symmetry ($\textrm{T}$), Space reversal ($ \textrm{P}$) and a combination of the $ \textrm{C},\textrm{P}\,\& \textrm{T}$-\textit{symmetries}. It is shown that these equations are Lorentz invariant, obey $ \textrm{C}$-\textit{symmetry} and that some violate $\textrm{T}$ and $\textrm{P}$-\textit{symmetry} while others do not and that they all obey $\textrm{PT}$-\textit{symmetry}. These symmetries show (or modestly said -- seem to suggest) that anti-particles have positive mass and energy but a negative rest-mass and the opposite sign in electronic charge. Through the inspection of these symmetries, a suggestion is (here) made to the effect that the rest-mass of a particle must be related to the electronic charge of that particle thus leading us to a possible resolution of whether or not Neutrinos do have a none-zero rest-mass. Additionally, we demonstrate that these equations have the potency to explain naturally the observed lepton generation phenomena. \\
\\
\textbf{{Keywords}:} Curved Spacetime, Gyromagnetic Ratio, Fundamental Particle, Symmetry.\\
\textbf{{PACS numbers (2009)}:} 03.65.Pm, 11.30.j, 04.62.b, 04.62.+v, 98.80.Jk, 04.40.b 
\linethickness{2pt}
\line(1,0){400}
\textsl{\begin{center}
\textbf{\footnotesize ``If one is working from the point of view of getting beauty in one's equation,\\ and if one has really sound insights, one is on a sure line of progress ...''}
\end{center}}
\begin{flushright}
-- \textbf{Paul Adrien Maurice Dirac} (1902-1984)
\end{flushright}
}
\end{abstract}

\maketitle

\small

\section{\sectionfont Introduction}

\PARstart{T}{hree} new Dirac-equivalent equations for a curved spacetime have been derived in the earlier reading Nyambuya ($2008$) and it has been shown  (\textit{in-principle}) that these equations have  the potency to naturally explain the non-zero anomalous gyromagnetic ratio of fermions without the aid of the elegant but rather un-natural Feynman diagrams as is the usual case when one sticks to the bare Dirac Theory. As part of the package of the shorting comings of the Dirac Equation ($1928a,b$) -- which is built on flat Minkowski spacetime --  \textit{i.e.}; in its natural and bare form, it is unable to account for the gyromagnetic ratio in excess of, or less than $2$.  The bare Dirac Theory needs some modification to explain these kind of observations and these modifications lead to the discovery of Quantum Electrodynamics (QED). 

Much to the glory of the Dirac Theory, it must be said that despite the shorting coming that it neatly and elegantly can only explain a gyromagnetic ratio $g=2$ which otherwise before its advent [the Dirac Equation], there was no theoretical explanation of why the gyromagnetic ratio of the Electron had a value of about 2. All that was known is that experiment demanded that it be about $2$ instead of $1$ as predicted from the then and only theory present to explain this -- \textit{i.e.}, the Schr\"odinger Theory of the atom. The ability to explain this $g=2$ gave the Dirac Equation its first initial success that lead to its quick acceptance. Further, the Dirac Theory was the first to naturally explain the origins of spin, which was only artificially inserted ``by hand'' into the equations of the Schr\"odinger Theory. 

In this reading, we continue to pedal, seeking more and more light that could link the three Curved Spacetime Dirac Equations (Nyambuya $2008$) with experience. Why three and not four, five equations or something? The reason is that spacetime has three kinds of curvatures, \textit{i.e.}, it can be quadratically, parabolically, or hyperbolically curved. Each of these curvature states has its own Curved Spacetime Dirac Equation to describe it. 

These three Curved Spacetime Dirac Equations  are given:

\begin{equation}
iA^{\mu}\gamma^{\mu}\partial_{\mu}\psi =\left(\frac{m_{0}c}{\hbar}\right)\psi,\label{dirac1}
\end{equation}

for the Quadratic Spacetime (QST);

\begin{equation}
iA^{\mu}\bar{\gamma}^{\mu}\partial_{\mu}\psi =\left(\frac{m_{0}c}{\hbar}\right)\psi,\label{dirac2}
\end{equation}

for the Hyperbolic  Spacetime (HST);

\begin{equation}
iA^{\mu}\hat{\gamma}^{\mu}\partial_{\mu}\psi =\left(\frac{m_{0}c}{\hbar}\right)\psi,\label{dirac3}
\end{equation}

for the Parabolic Spacetime (PST); where $\hbar=1.06\times10^{-35}\textrm{Js}$ is the normalized Planck constant, $c=2.99792458\times10^{8}\textrm{ms}^{-1}$ is the sacrosanct speed of light, $m_{0}$ the rest-mass of the particle in question and, $\psi$ is the Dirac four spinor of that particle and this is given by:

\begin{equation}
\psi=\left(\begin{array}{c}
\psi_{0}\\
\psi_{1}\\
\psi_{2}\\
\psi_{3}\end{array}\right)\label{4spinor},
\end{equation}

and  $\gamma^{\mu}$, $\bar{\gamma}^{\mu}$ and $\hat{\gamma}^{\mu}$ are the $4\times4$ Dirac gamma, gamma-bar  and the gamma-hat matrices respectively and  these are given by:

\begin{equation}
\begin{array}{c c}
\gamma^{0}=
\left(\begin{array}{c c}
\textbf{I} & \mathbf{0}\\
\mathbf{0} & -\textbf{I} \\
\end{array}\right)
,\,\,\,\,
\gamma^{i}=
\left(\begin{array}{c c}
\textbf{0} & \boldsymbol{\sigma}^{i}\\
-\boldsymbol{\sigma}^{i} & \textbf{0} \\
\end{array}\right),
\end{array}\label{gamma-m}
\end{equation}

\begin{equation}
\begin{array}{c c}
\bar{\gamma}^{0}=
\left(\begin{array}{c c}
\textbf{I} & \mathbf{0}\\
\mathbf{0} & -\textbf{I} \\
\end{array}\right)
,\,\,\,\,
\bar{\gamma}^{i}=
\frac{1}{2}\left(\begin{array}{c c}
2\textbf{I} & i\sqrt{2}\boldsymbol{\sigma}^{i}\\
-i\sqrt{2}\boldsymbol{\sigma}^{i} & -2\textbf{I} \\
\end{array}\right),
\end{array}\label{gamma-bar-m}
\end{equation}

\begin{equation}
\begin{array}{c c}
\hat{\gamma}^{0}=
\pm\left(\begin{array}{c c}
\textbf{I} & \mathbf{0}\\
\mathbf{0} & -\textbf{I} \\
\end{array}\right)
,\,\,\,\,
\hat{\gamma}^{i}=
\mp\frac{1}{2}\left(\begin{array}{c c}
2\textbf{I} & i\sqrt{2}\boldsymbol{\sigma}^{i}\\
-i\sqrt{2}\boldsymbol{\sigma}^{i} & -2\textbf{I} \\
\end{array}\right),
\end{array}\label{gamma-hat-m}
\end{equation}

where $\boldsymbol{\sigma}^{i}$ are the $2\times2$ Pauli matrices, $\textrm{\textbf{0}}$ is the $2\times2$ null matrix  and $\textrm{\textbf{I}}$ the $2\times2$ identity matrix. The Greek indices \textit{e.g.}, $\mu$ -- shall; unless otherwise specified, be understood to run from $0$ to $3$, \textit{i.e.} $\mu=0,1,2,3$.  

Now, to compatify the there equations, if we set $\gamma^{\mu}_{(1)}=\gamma^{\mu}$, $\gamma^{\mu}_{(2)}=\bar{\gamma}^{\mu}$ and $\gamma^{\mu}_{(3)}=\hat{\gamma}^{\mu}$, then  equations (\ref{dirac1}), (\ref{dirac2}) and (\ref{dirac3}) can be written compactly as:

\begin{equation}
iA^{\mu}\gamma^{\mu}_{(a)}\partial_{\mu}\psi=\left(\frac{m_{0}c}{\hbar}\right)\psi,\label{dirac4}
\end{equation}

where $a=1,2,3$. We have put the $a$-index inside the brackets to indicate that it is not an active index like $\mu$; we shall employ this notation to all non-active indices. 

Before we proceed, we need to clear the air as some may misread the product $A^{\mu}\gamma^{\mu}_{(a)}$ because of the double $\mu$-index. This product is a vector and must be viewed as one composite unit -- \textit{i.e.}, $\Gamma^{\mu}_{(a)}=A^{\mu}\gamma^{\mu}_{(a)}$. For example, taking the case $a=1$, which gives the usual Dirac matrices -- \textit{i.e.}, $\Gamma^{\mu}_{(1)}=A^{\mu}\gamma^{\mu}_{(1)}=A^{\mu}\gamma^{\mu}$, and writing it in full, we will have:

\begin{equation}
\begin{array}{c c}

\Gamma^{0}_{(1)}=A^{0}\left(\begin{array}{r r r r}
1 & 0 & 0 & 0 \\
0 & 1 & 0 & 0\\
0 & 0 & -1 & 0\\
0 & 0 & 0 & -1
\end{array}\right), &

\Gamma^{1}_{(1)}=A^{1}\left(\begin{array}{r r r r}
0 & 0 & 0 & 1 \\
1 & 0 & 1 & 0\\
0 & -1 & 0 & 0\\
-1 & 1 & 0 & 0
\end{array}\right), \\
\\
\Gamma^{2}_{(1)}=
A^{2}\left(\begin{array}{r r r r}
0 & 0 & 0 & -i\\
0 & 0 & i & 0\\
0 & i & 0 & 0\\
-i & 0 & 0 & 0
\end{array}\right), &

\Gamma^{3}_{(1)}=
A^{3}\left(\begin{array}{rrrr}
0 & 0 & 1 & 0\\
0 & 0 & 0 & -1\\
-1 & 0 & 0 & 0\\
0 & 1 & 0 & 0
\end{array}\right) 

\end{array}.
\end{equation}

Written in this form, it is clear that the matrices are constants while the object $A^{\mu}$ is a real vector (since $\mu=0,1,2,3$ there are four components of the vector $A^{\mu}$). The index $\mu$ appearing in the $4\times4$ matrices $\gamma^{\mu}_{(a)}$ may deceive one to a point that they may view it as a vector, while in actual fact, that is not the case. Thus equations (\ref{dirac1}), (\ref{dirac2}) and (\ref{dirac3}) (or effectively equation \ref{dirac4}) can then be written more compactly as:

\begin{equation}
i\Gamma^{\mu}_{(a)}\partial_{\mu}\psi=\left(\frac{m_{0}c}{\hbar}\right)\psi,\label{dirac5}
\end{equation}

which is much more clear to visualize its Lorentz invariance. We shall stress that we prefer the form (\ref{dirac4}) to the form (\ref{dirac5}) because (\ref{dirac4}) makes clear the existence of the vector $A^{\mu}$ while (\ref{dirac5}) sort of conceals it and one always has to be conscious of its existence. Also, symbolically and more so visually, this distinguishes these equations from the orginal Dirac Equation.

These equations, \textit{i.e.} (\ref{dirac4}); have (\textit{in principle}) been shown in the reading Nyambuya ($2008$) to be able to explain the anomalous gyromagnetic ratio of fermions and the present reading  adds on by identifying the vector quantity $A^{\mu}$ that was therein not identified, \textit{vis} its physical meaning and significance.

\textbf{{Definition}:} \textit{The vector $A^{\mu}$ which has (in principle) been shown in Nyambuya ($2007$) to be able to represent the Electromagnetic, the Weak, and the Strong force shall -- for the sake of keeping this reading as simple as one can; (here) be identified with the Electromagnetic field of the particle in question.}

In the reading Nyambuya ($2007$), the function $A^{\mu}$ is dealt with and (there) it is shown to be able to represent the known three forces of nature, \textit{i.e.}, the Electromagnetic, the Weak, and the Strong force. We here direct the reader to this reading for a better and clear understanding of this and if for some reason the reader disagrees with these ideas since these ideas are still in their infancy and further work on them is underway; the reader is here persuaded to accept the definition of this vector as given above.

Regarding equation (\ref{dirac4}), it should be said that from a mathematical standpoint, there is no reason to believe in the non-validity of this equation as it legitimately flows from the nature of a curved spacetime with the only controversy (if any) being the identification of $A^{\mu}$ with the Electromagnetic field of the particle. On physical grounds, yes it can be rejected if it does not conform with experience. This said, it appears to me, one sure way to seek more and further ground for this equation is the obvious, apply it say, to the Hydrogen atom and search for any anomalous solutions and check if these anomalous solutions lead to any observed anomalous phenomena associated with the Hydrogen atom. Given the non-linear nature of the equations, one will most certainly need to solve the equations numerically and also one will have to seek the field $A^{\mu}$ from outside the provinces of the present theory, they will have to do so from the Unified Field Theory (Nyambuya $2007$)  -- we are not going to do this here!

Further, an advantage of this equation, which appears so clear to me, is that, one does not and will not need the many Feynman diagrams to calculate the anomalous gyromagnetic ratio (\textit{e.g.} see  Brodsky \textit{et al.} 2004; Knecht 2002; Laporta \& Remiddi 1996; Karplus \& Kroll 1950) as this equation clearly predicts a deviation from $g=2$ as a direct consequence of the fact that spacetime is curved and that  particles do have a finite spatial size and are not to be treated as condensed point-sources.

From hereon, \textit{i.e.}, after the present introductory section; the rest of this reading is subdivided into six sections. In the succeeding section, \textit{i.e.}, \S (\ref{neweqn}), we increase the number of Dirac Curved Spacetime Equations from $3$ to $192$ -- this is certainly not nice as it appears to increase the complexity. \textit{Vis} the aforesaid, it should be said that consistent with the way, manner  or method in which equations (\ref{dirac1}, \ref{dirac2} \& \ref{dirac3}) have been derived, the additional equations are equations that are mathematically permitted to exist and we bring these in, for nothing other than the quest for completeness. The three family structure of the Dirac Curved Spacetime Equations exhibited by the original three equations (\ref{dirac1}, \ref{dirac2} \& \ref{dirac3}) is not destroyed but maintained -- this in great many ways returns the theory to its original simplicity as one can understand the additional equations by simple understanding the three original equations. 

In \S (\ref{props}) the properties of the Curved Spacetime Dirac Equations are formally laid down and these are the Lorentz Symmetry, $ \textrm{C}$, $ \textrm{P}$, $ \textrm{T}$ and a combinations of the $ \textrm{C,P\,\&T}$-symmetries ($ \textrm{C}$ for charge conjugation symmetry, $\textrm{P}$ for space reversal symmetry or parity and $ \textrm{T}$ for time reversal symmetry). In \S (\ref{cpv}) special attention is paid to the CP-symmetry violation. In \S (\ref{lgp}) we show that the Curved Spacetime Dirac Equations do \textit{predict}  a three family structure of particles that are marked by their masses just as is seen in leptons -- this hints at a solution to the generation problem seen to be exhibited by leptons. If we did not know better, we would seize this moment and propose that this must explain the lepton generation problem. We not not do that but simple point out that this hints at a solution. Only in a future reading (Nyambuya $2009$) are we going to this. In \S (\ref{netrino}) we investigate (\textit{vis} the fact that neutrinos are leptons with zero electronic charge) the implied finding that the rest mass of a particle is related to the electronic charge. Lastly, in \S (\ref{dis}) we give an overall discussion and lay down conclusion(s) that can be drawn from this reading.

\section{\sectionfont New more Equations\label{neweqn}}

To the three equations [\ref{dirac1}, \ref{dirac2} and \ref{dirac3}], we shall add $189$ more equations by noting that:

\begin{enumerate}
\renewcommand{\theenumi}{(\arabic{enumi})}
\item$\,$ The bare Dirac Equation, $i\gamma^{\mu}\partial_{\mu}\psi=(m_{0}c/\hbar)\psi$, is known to satisfy the equation, $\eta^{\mu\nu}\partial_{\mu}\partial_{\nu}\psi=(m_{0}c/\hbar)^{2}\psi$ ($\eta^{\mu\nu}$ is the flat Minkowski metric of spacetime), which in actual fact is the Klein-Gordon Equation and this can be generalized for a curved spacetime to:

\begin{equation}
g^{\mu\nu}\partial_{\mu}\partial_{\nu}\psi=\left(\frac{m_{0}c}{\hbar}\right)^{2}\psi.\label{Klein-Gordon 1}
\end{equation}

If we modified the Dirac Equation so that it reads, $i\gamma^{\mu}\partial_{\mu}\psi~=~(m_{0}c/\hbar)\tilde{\gamma}^{\ell}\psi$, where $\tilde{\gamma}^{\ell\dagger}\tilde{\gamma}^{\ell}=\textrm{I}$, and $\tilde{\gamma}^{\ell}$ are some $4\times4$ matrices and $\textrm{I}$ is the $4\times4$ identity matrix, we would -- from this modified equation; arrive at equation (\ref{Klein-Gordon 1}) \textit{via} the same path one arrives at this equation from the bare Dirac Equation. This modification adds $15$ more equations to the already existing bare Dirac Equation because there are sixteen $\tilde{\gamma}^{\ell}$ matrices that satisfy the condition $\tilde{\gamma}^{\ell\dagger}\tilde{\gamma}^{\ell}=\textrm{I}$ -- it will be demonstrated shortly that, there are indeed $16$ such matrices $\tilde{\gamma}^{\ell}$ that meet the condition $\tilde{\gamma}^{a\dagger}\tilde{\gamma}^{\ell}=\textrm{I}$.  
\\
\item$\,$ The bare Dirac Equation, can be shown to satisfy the equation, $\eta^{\mu\nu}\partial_{\mu}\psi^{\dagger}\partial_{\nu}\psi=(m_{0}c/\hbar)^{2}\psi^{\dagger}\psi$ and one arrives at this by multiplying the Dirac Equation from the left by its complex conjugate. This can be generalized for a curved spacetime to:

\begin{equation}
g^{\mu\nu}\left(\partial_{\mu}\psi^{\dagger}\partial_{\nu}\psi\right) = \left(\frac{m_{0}c}{\hbar}\right)^{2}\psi^{\dagger}\psi.\label{jjeqn0}
\end{equation}

\end{enumerate}

As has been shown in Nyambuya ($2008$), the metric, $g^{\mu\nu}$, can be written in terms of $A^{\mu}$, \textit{i.e.,} $g^{\mu\nu}=g^{\mu\nu}(A^{\mu}A^{\nu})$, and in this form, it takes three different forms. So what we shall do here is to seek \textit{\textbf{all the equations}}  in-terms of $A^{\mu}$ and $\psi$ that satisfy the generalized equations (\ref{Klein-Gordon 1}) and (\ref{jjeqn0}). Before doing so, let us address the issue of the metric tensor and its relation to the gravitational field.

\textbf{{The Metric Tensor and the Gravitational Phenomena}:} At this point one may wonder if the metric $g^{\mu\nu}$ is the conventional metric of spacetime as is the case in the General Theory \textit{of} Relativity (GTR) since (as per the earlier definition) it is now a function of the Electromagnetic vector potential meaning to say it now represents the Electromagnetic phenomena and not gravitation as is the usual case in the GTR. As has been said already that, in the reading Nyambuya ($2008$), this vector, $A^{\mu}$, is identified with the Electromagnetic field, the Weak and the Strong force but here we choose to investigate only the Electromagnetic phenomena emanating from this vector.  This reading Nyambuya ($2007$), proposes a new theory that tries to tie together all the forces of \textit{Nature} and as has been said, further work is in progress on furthering these ideas. There in (Nyambuya $2007$), the gravitational field is part of the metric but is described by a scaler potential, \textit{i.e.}, $g^{\mu\nu}_{(a)}=\frac{1}{2}\varphi\left\{\gamma^{\mu}_{(a)},\gamma^{\mu}_{(a)}\right\}A^{\mu}A^{\nu}$ where $\varphi$ is the scaler that describes gravitation -- for better clarity, we direct the reader to Nyambuya ($2007$).

Again, before proceeding, we would like to take this time to caution the reader that in order to follow smoothly the flow of this reading, they ought to be prepared to do some tedious algebra in-order to verify and satisfy themselves of the correctness of the equations presented herein, because we do not go through all the steps leading to the final equations. We, however have taken the liberty to spell out the steps we have taken to reach whatever equations that we have presented. Otherwise, to put all the mathematical steps, would reduce this reading into an unnecessary litter and nightmare of symbols that is not appealing to the naked eye. Additionally, we strongly encourage the reader to at least have the readings Nyambuya ($2007,2008$)  with them if they hope to make sense of the present reading as the present reading is intimately tied to these readings. 

Now proceeding,  we present the three cases of the three forms of the metric $g^{\mu\nu}$ and these arise because spacetime is either Quadratically, Hyperbolically or Parabolically curved.

\textbf{(1) \underline{Quadratic Spacetime}:} The first form is for  the quadratic spacetime whose metric has zero off-diagonal terms. For a $\textit{3}$-Dimensional Space, this space has the shape of a football -- albeit a bumpy one with no smooth surface if $A^{\mu}$ is not a constant but a function of the spacetime coordinates $x^{\mu}$. The metric of this spacetime is given by:

\begin{equation}
[g^{\mu\nu}]=\left(
\begin{array}{cccc}
A^{0}A^{0} & 0 & 0 &0\\
0 & -A^{1}A^{1} & 0 &0\\
0 & 0 & -A^{2}A^{2} & 0\\
0 & 0 & 0 & -A^{3}A^{3}
\end{array}\right),\label{fmetric}
\end{equation}

where $A^{\mu}$ is (as has been discussed) the four vector potential which represents the Electromagnetic field of the particle.  In much the same manner as has been shown in Nyambuya ($2008$),  where the three Curved Spacetime Dirac Equations where first derived, equation (\ref{Klein-Gordon 1}) has, for the QST setting, the solutions:

\begin{equation}
iA^{\mu}\gamma^{\mu}\partial_{\mu}\psi =\left(-1\right)^{\ell}\left(\frac{m_{0}c}{\hbar}\right)\tilde{\gamma}^{\ell}\psi,\label{diracf1}
\end{equation}

where $\ell=0,1,2, ...,15$ and:

\begin{equation}
iA^{\mu}\gamma^{\mu}\partial_{\mu}\psi =-\left(-1\right)^{\ell}\left(\frac{m_{0}c}{\hbar}\right)i\tilde{\gamma}^{\ell}\psi,\label{diracf2}
\end{equation}

where  the term $\left(-1\right)^{\ell}$  has been carefully inserted so that these equation obey charge conjugation symmetry -- this will become clear in \S (\ref{cconj}). The  matrices $\tilde{\gamma}^{\ell}$ (lets call them the gamma-tilde matrices) as already said; these are  sixteen  $4\times4$ matrices such that $\tilde{\gamma}^{\ell\dagger}\tilde{\gamma}^{\ell}=\textrm{I}$  and these sixteen matrices are ($\gamma^{\mu}, \textrm{I},\gamma^{5},\sigma^{\mu\nu},\gamma^{\mu}\gamma^{5}$) where $\sigma^{\mu\nu}=\gamma^{\mu}\gamma^{\nu}-\gamma^{\nu}\gamma^{\mu}$ and $\gamma^{5}=i\gamma^{0}\gamma^{1}\gamma^{2}\gamma^{3}$. Written in full, these matrices are given by (\ref{gmatrices1}) and the fifth column in (\ref{gmatrices1}) gives the compact form of the particular row. 

\begin{widetext}
\centering

\begin{equation}
\begin{array}{|r | r | r | r ||l|}
\hline
\multicolumn{4}{|c||}{\textbf{\tabletitlefont The Tilde-matrices}
} & \textbf{\tabletitlefont Condensed\,Form}\\
\hline
\tilde{\gamma}^{0}= \left(\begin{array}{cc}
\textrm{\textbf{I}} & \textbf{0} \\
\textbf{0} & -\textrm{\textbf{I}} \\
\end{array}\right) & 

\tilde{\gamma}^{1} = \left(\begin{array}{cc}
\textbf{0}  & \boldsymbol{\sigma}^{1} \\
-\boldsymbol{\sigma}^{1} & \textbf{0} \\
\end{array}\right) & 

\tilde{\gamma}^{2} = i\left(\begin{array}{cc}
\textbf{0}  & \boldsymbol{\sigma}^{2} \\
-\boldsymbol{\sigma}^{2} & \textbf{0} \\
\end{array}\right) & 

\tilde{\gamma}^{3} = \left(\begin{array}{cc}
\textbf{0}  & \boldsymbol{\sigma}^{3} \\
-\boldsymbol{\sigma}^{3} & \textbf{0} \\
\end{array}\right) 
&

\rightrightarrows\gamma^{\mu}_{*}\\
\hline
\tilde{\gamma}^{4} = \left(\begin{array}{cc}
 \textrm{\textbf{I}} & \textbf{0}  \\
 \textbf{0} & \textrm{\textbf{I}}  \\
\end{array}\right) 

& 
 
\tilde{\gamma}^{5} = \left(\begin{array}{cc}
\textbf{0}  & \boldsymbol{\sigma}^{1} \\
\boldsymbol{\sigma}^{1} & \textbf{0} \\
\end{array}\right)
&

\tilde{\gamma}^{6} = i\left(\begin{array}{cc}
\textbf{0}  & \boldsymbol{\sigma}^{2} \\
\boldsymbol{\sigma}^{2} & \textbf{0} \\
\end{array}\right)  

 & 

\tilde{\gamma}^{7}= \left(\begin{array}{cc}
\textbf{0}  & \boldsymbol{\sigma}^{3} \\
\boldsymbol{\sigma}^{3} & \textbf{0} \\
\end{array}\right) 

&
\rightrightarrows\gamma^{0}\gamma^{\mu}_{*}\\
\hline
\tilde{\gamma}^{8}= \left(\begin{array}{cc}
\textbf{0} &\textrm{\textbf{I}}  \\
 -\textrm{\textbf{I}} & \textbf{0} \\
\end{array}\right) 

&

\tilde{\gamma}^{9} = \left(\begin{array}{c c}
\boldsymbol{\sigma}^{1} & \textbf{0} \\
\textbf{0} & -\boldsymbol{\sigma}^{1} \\
\end{array}\right) 

&

\tilde{\gamma}^{10}= i\left(\begin{array}{ c c}
\boldsymbol{\sigma}^{2} & \textbf{0} \\
\textbf{0} & -\boldsymbol{\sigma}^{2} \\
\end{array}\right)
&
\tilde{\gamma}^{11} = \left(\begin{array}{c c}
\boldsymbol{\sigma}^{3} & \textbf{0} \\
\textbf{0} & -\boldsymbol{\sigma}^{3} \\
\end{array}\right)

&

\rightrightarrows\gamma^{\mu}_{*}\gamma^{5}\\
\hline
\tilde{\gamma}^{12} = \left(\begin{array}{rr}
 \textbf{0} & \textrm{\textbf{I}} \\
\textrm{\textbf{I}} & \textbf{0} \\
\end{array}\right) 
 & 
\tilde{\gamma}^{13} = \left(\begin{array}{l l}
\boldsymbol{\sigma}^{1} & \textbf{0} \\
\textbf{0} & \boldsymbol{\sigma}^{1} \\
\end{array}\right) &

\tilde{\gamma}^{14}= i\left(\begin{array}{ l l}
\boldsymbol{\sigma}^{2} & \textbf{0} \\
\textbf{0} & \boldsymbol{\sigma}^{2} \\
\end{array}\right) & 

\tilde{\gamma}^{15} = \left(\begin{array}{l l}
\boldsymbol{\sigma}^{3} & \textbf{0} \\
\textbf{0} & \boldsymbol{\sigma}^{3} \\
\end{array}\right)  
 &
\rightrightarrows\gamma^{0}\gamma^{\mu}_{*}\gamma^{5}\\
\hline
\end{array}. 
\label{gmatrices1}
\end{equation}
\end{widetext}

The matrices $\tilde{\gamma}^{\ell}$ are pure constants (real and not complex) with no dependence whatsoever on the frame like in the Dirac Theory, where the $\gamma^{\mu}$-matrices have a frame dependence -- this is not the case here. We shall show in \S (\ref{linv})  that the $\gamma^{\mu}_{(a)}$-matrices (which very shortly will be transformed to become the $\gamma^{\mu}_{(as)}$-matrices) share this property as-well. 

Proceeding \textbf{...} we note that equation (\ref{jjeqn0}) can be written in a different but equivalent from as:

\begin{equation}
g^{\mu\nu}\left(\partial_{\mu}\psi^{\dagger}\partial_{\nu}\psi\right) = \left(\frac{m_{0}c}{\hbar}\right)^{2}\psi^{T}\psi^{*},\label{jjeqn}
\end{equation}

where the superscript $T$ and $*$ are the transpose and complete conjugate  operations on the wavefunction $\psi$. From this, we will have two more equations, \textit{i.e.} \textbf{(1)}:

\begin{equation}
iA^{\mu}\gamma^{\mu}\partial_{\mu}\psi =\left(-1\right)^{\ell}\left(\frac{m_{0}c}{\hbar}\right)\tilde{\gamma}^{\ell}\psi_{c},\label{ndiracc1}
\end{equation}

where $\psi_{c}=\gamma^{0}\gamma^{2}\psi^{*}$ and the matrices $\gamma^{0}$ and $\gamma^{2}$ have been chosen carefully so that these equations obey charge conjugation symmetry (this will be seen in \S \ref{cconj}). \textbf{(2}) The other equation satisfying equation (\ref{jjeqn}) is:

\begin{equation}
iA^{\mu}\gamma^{\mu}\partial_{\mu}\psi =-\left(-1\right)^{\ell}\left(\frac{m_{0}c}{\hbar}\right)i\tilde{\gamma}^{\ell}\psi_{c}.\label{ndiracc2}
\end{equation}

Equations (\ref{diracf1}), (\ref{diracf2}), (\ref{ndiracc1}) and (\ref{ndiracc2}) shall hereafter be referred to as the Quadratic Spacetime Equations (QST-Eqns).

\textbf{(2) \underline{Hyperbolic Spacetime}:} In the third form, $g^{\mu\nu}$ is representative of a ``positively'' ($\lambda=+1$, this will be defined later in \S \ref{invrestmass}) curved spacetime, which for a $\textit{3}$-Dimensional Space it has the shape of a hyperboloid (\textit{e.g.}, like a saddle used on horse-back). The metric of this spacetime is given by:

\begin{equation}
[g^{\mu\nu}]=\left(
\begin{array}{r r r r}
A^{0}A^{0} & A^{0}A^{1} & A^{0}A^{2} & A^{0}A^{3}\\
A^{1}A^{0} & -A^{1}A^{1} & A^{1}A^{2} & A^{1}A^{3}\\
A^{2}A^{0} & A^{2}A^{1} & -A^{2}A^{2} & A^{2}A^{3}\\
A^{3}A^{0} & A^{3}A^{1} & A^{3}A^{2} & -A^{3}A^{3}
\end{array}\right),
\end{equation}

and  just as the case for the QST, this form of the metric will result in four equations satisfying (\ref{Klein-Gordon 1}) and (\ref{jjeqn}), \textit{i.e.}:

\begin{equation}
iA^{\mu}\bar{\gamma}^{\mu}\partial_{\mu}\psi =\left(-1\right)^{\ell}\left(\frac{m_{0}c}{\hbar}\right)\tilde{\gamma}^{\ell}\psi\label{pdirac1},
\end{equation}

\begin{equation}
iA^{\mu}\bar{\gamma}^{\mu}\partial_{\mu}\psi =-\left(-1\right)^{\ell}\left(\frac{m_{0}c}{\hbar}\right)i\tilde{\gamma}^{\ell}\psi\label{pdirac2},
\end{equation}

(these two equations satisfy \ref{Klein-Gordon 1}), and; 

\begin{equation}
iA^{\mu}\bar{\gamma}^{\mu}\partial_{\mu}\psi =\left(-1\right)^{\ell}\left(\frac{m_{0}c}{\hbar}\right)\tilde{\gamma}^{\ell}\psi_{c},\label{npdirac1}
\end{equation}

\begin{equation}
iA^{\mu}\bar{\gamma}^{\mu}\partial_{\mu}\psi =-\left(-1\right)^{\ell}\left(\frac{m_{0}c}{\hbar}\right)i\tilde{\gamma}^{\ell}\psi_{c}.\label{npdirac2}
\end{equation} 

(these two equations satisfy \ref{jjeqn}). In-passing, it should be said that, it is important and required of a UFT such as Nyambuya ($2008$) to contain in it the Dirac Equation  since it is an equation describing fundamental particles, thus the fact that this is so for this UFT, is appealing with regard to it [the present theory] containing an element or a grain of truth. 

\textbf{(3) \underline{Parabolic Spacetime}:} In the second form, $g^{\mu\nu}$ is representative of a ``negatively'' ($\lambda=-1$, this will be defined later in \S \ref{invrestmass}) curved spacetime which for a $\textit{3}$-Dimensional Space it has the shape of a paraboloid (\textit{e.g.}, like a rugby ball or an egg). The metric of this spacetime is given by:

\begin{equation}
[g^{\mu\nu}]=\left(
\begin{array}{r r r r}
A^{0}A^{0} & -A^{0}A^{1} & -A^{0}A^{2} & -A^{0}A^{3}\\
-A^{1}A^{0} & -A^{1}A^{1} & -A^{1}A^{2} & -A^{1}A^{3}\\
-A^{2}A^{0} & -A^{2}A^{1} & -A^{2}A^{2} & -A^{2}A^{3}\\
-A^{3}A^{0} & -A^{3}A^{1} & -A^{3}A^{2} & -A^{3}A^{3}
\end{array}\right),
\end{equation}
\\
and  just as the case for ``flat'' and ``positively'' curved spacetime, this form of the metric will result in four equations satisfying (\ref{Klein-Gordon 1}) and (\ref{jjeqn}), \textit{i.e.}:

\begin{equation}
iA^{\mu}\hat{\gamma}^{\mu}\partial_{\mu}\psi =\left(-1\right)^{\ell}\left(\frac{m_{0}c}{\hbar}\right)\tilde{\gamma}^{\ell}\psi,\label{ndirac1}
\end{equation}

\begin{equation}
iA^{\mu}\hat{\gamma}^{\mu}\partial_{\mu}\psi =-\left(-1\right)^{\ell}\left(\frac{m_{0}c}{\hbar}\right)i\tilde{\gamma}^{\ell}\psi,\label{ndirac2}
\end{equation}

(these two equations satisfy \ref{Klein-Gordon 1}), and; 

\begin{equation}
iA^{\mu}\hat{\gamma}^{\mu}\partial_{\mu}\psi =\left(-1\right)^{\ell}\left(\frac{m_{0}c}{\hbar}\right)\tilde{\gamma}^{\ell}\psi_{c},\label{nndirac1}
\end{equation}

\begin{equation}
iA^{\mu}\hat{\gamma}^{\mu}\partial_{\mu}\psi =-\left(-1\right)^{\ell}\left(\frac{m_{0}c}{\hbar}\right)i\tilde{\gamma}^{\ell}\psi_{c}.\label{nndirac2}
\end{equation} 

(these two equations satisfy \ref{jjeqn}). Effectively, equations (\ref{ndirac1}), (\ref{ndirac2}), (\ref{nndirac1}), and (\ref{nndirac2}) are similar but different and distinct equations from equations (\ref{pdirac1}), (\ref{pdirac2}), (\ref{npdirac1}), and (\ref{npdirac2}) respectively in that their energies are flipped. That is to say, if for the equations (\ref{ndirac1}), (\ref{ndirac2}), (\ref{nndirac1}), and (\ref{nndirac2}) the energy solutions are the ordered pair $\left<E_{+},E_{-}\right>$ where $E_{+}>0$ is the positive energy solution and $E_{-}<0$ is the negative energy solution, then for  equations (\ref{pdirac1}), (\ref{pdirac2}), (\ref{npdirac1}), and (\ref{npdirac2}), the energy solutions are $\left<|E_{-}|,-E_{+}\right>$ respectively.

In total, we have a set of twelve equations with each set having sixteen equations! Let us collect these and write them in a neat tabular form as shown in table (\ref{equations}). In this table the particle equations have been arranged in families and hierarchies. We have the Majorana-Type particles, which are the particles described by the equations in the first and second row. The name Majorana comes in because these equations have the Majorana form, \textit{i.e.}, with the Dirac spinor-$\psi$ on the left hand side of the equation and the Majorana spinor-$\psi_{c}$ on the right hand side; and the Majorana-Type II for which $\lambda=0$, $|A^{\mu}|=1$ and $\ell=4$ gives the true Majorana equation (see Majorana $1934$). Also, the name Dirac Family (third and fourth row) comes in because these equations for the case $\lambda=0$, $|A^{\mu}|=1$ and $\ell=4$, we have the bare Dirac Equation. Also, on both sides of the equation, we have just the Dirac spinor-$\psi$, this is the same form as the Dirac Equation.

Equations (\ref{ndirac1}), (\ref{ndirac2}), (\ref{nndirac1}),  and (\ref{nndirac2}) shall hereafter be referred to as the Parabolic Spacetime Equations (PST-Eqns) while equations (\ref{pdirac1}), (\ref{pdirac2}), (\ref{npdirac1}), and (\ref{npdirac2}) shall be referred to as the Hyperbolic Spacetime Equations (HST-Eqns). Together as a collective, equations (\ref{ndirac1}), (\ref{ndirac2}), (\ref{nndirac1}), (\ref{nndirac2}), (\ref{pdirac1}), (\ref{pdirac2}), (\ref{npdirac1}), and (\ref{npdirac2}) shall simple be referred to as the Diagonally Curved Spacetime Equations (DCST-Eqns), the term ``Diagonal'' comes in here because the PST and HST have non-zero off diagonals terms in their metric.

As will be shown in \S (\ref{invrestmass}), considering only the positive energy solutions of these equations, the particles for which $\lambda=0$ are expected to be the least massive while the particles for which  $\lambda=+1$ are expected to be the most massive with those particles for which  $\lambda=-1$, their mass will lay in the intermediate range and thus the mass hierarchy problem exhibited by fermions naturally finds a possible explanation, hence the last row with the hierarchies labels HI, HII and HIII; these are in-terms of the mass of the particles.

\begin{widetext}

\begin{table}[!h]
\centering
\caption{\tabletitlefont Collected Equations}\label{equations}
\vspace*{0.3cm}
{\tabletextfont
\begin{tabular}{ | l || l | l | l |}

\hline
\multicolumn{1}{|l||}{\textbf{Family}} & \multicolumn{3}{|c|}{\textbf{Nature \textit{of} Spacetime}} \\
 \hline
\multicolumn{1}{|c||}{$\downarrow\,\,\,\,\,\,\,\,\,\,\,\,\,\,\,\,\,\,$}  & \multicolumn{1}{|c||}{$\lambda=0$} & \multicolumn{1}{|c||}{$\lambda=-1$} &  \multicolumn{1}{|c||}{$\lambda=+1$} \\
\hline
\multicolumn{1}{|r||}{$\downarrow$ \textbf{Energy}$\mapsto$ }  & \multicolumn{1}{|c||}{$\left\langle E_{+}^{(0)}, E_{-}^{(0)}\right\rangle$ or $\left\langle E_{-}^{(0)}, E_{+}^{(0)}\right\rangle$} & \multicolumn{1}{|c||}{$\left\langle E_{+}^{(-)}, E_{-}^{(-1)}\right\rangle$} &  \multicolumn{1}{|c||}{$\left\langle E_{+}^{(+1)}, E_{-}^{(+1)}\right\rangle$ } \\
\hline\hline
Majorana Type I &  $iA^{\mu}\gamma^{\mu}\partial_{\mu}\psi =\,\,\,\,\,\left(-1\right)^{\ell}\left(\frac{m_{0}c}{\hbar}\right)\tilde{\gamma}^{\ell}\psi_{c}$             & $iA^{\mu}\bar{\gamma}^{\mu}\partial_{\mu}\psi =\,\,\,\,\,\left(-1\right)^{\ell}\left(\frac{m_{0}c}{\hbar}\right)\tilde{\gamma}^{\ell}\psi_{c}$  &   $iA^{\mu}\bar{\gamma}^{\mu}\partial_{\mu}\psi =\,\,\,\,\,\left(-1\right)^{\ell}\left(\frac{m_{0}c}{\hbar}\right)\hat{\gamma}^{\ell}\psi_{c}$  \\
\hline
Majorana Type II & $iA^{\mu}\gamma^{\mu}\partial_{\mu}\psi =-\left(-1\right)^{\ell}\left(\frac{m_{0}c}{\hbar}\right)i\tilde{\gamma}^{\ell}\psi_{c}$               & $iA^{\mu}\bar{\gamma}^{\mu}\partial_{\mu}\psi =-\left(-1\right)^{\ell}\left(\frac{m_{0}c}{\hbar}\right)i\tilde{\gamma}^{\ell}\psi_{c}$ &   $iA^{\mu}\bar{\gamma}^{\mu}\partial_{\mu}\psi =-\left(-1\right)^{\ell}\left(\frac{m_{0}c}{\hbar}\right)i\hat{\gamma}^{\ell}\psi_{c}$  \\
\hline
Dirac Type I   & $iA^{\mu}\gamma^{\mu}\partial_{\mu}\psi =\,\,\,\,\,\left(-1\right)^{\ell}\left(\frac{m_{0}c}{\hbar}\right)\tilde{\gamma}^{\ell}\psi$             &  $iA^{\mu}\bar{\gamma}^{\mu}\partial_{\mu}\psi =\,\,\,\,\,\left(-1\right)^{\ell}\left(\frac{m_{0}c}{\hbar}\right)\tilde{\gamma}^{\ell}\psi$     &  $iA^{\mu}\bar{\gamma}^{\mu}\partial_{\mu}\psi =\,\,\,\,\,\left(-1\right)^{\ell}\left(\frac{m_{0}c}{\hbar}\right)\hat{\gamma}^{\ell}\psi$   \\
\hline
Dirac Type II  & $iA^{\mu}\gamma^{\mu}\partial_{\mu}\psi =-\left(-1\right)^{\ell}\left(\frac{m_{0}c}{\hbar}\right)i\tilde{\gamma}^{\ell}\psi$              &   $iA^{\mu}\bar{\gamma}^{\mu}\partial_{\mu}\psi =-\left(-1\right)^{\ell}\left(\frac{m_{0}c}{\hbar}\right)i\tilde{\gamma}^{\ell}\psi$ &  $iA^{\mu}\bar{\gamma}^{\mu}\partial_{\mu}\psi =-\left(-1\right)^{\ell}\left(\frac{m_{0}c}{\hbar}\right)i\hat{\gamma}^{\ell}\psi$    \\
\hline\hline
\textbf{Hierarchy}$\,\,\longmapsto$ & \multicolumn{1}{|c||}{$\textrm{I}$} & \multicolumn{1}{|c||}{$\textrm{II}$}  & \multicolumn{1}{|c||}{$\textrm{III}$}\\
\hline
\end{tabular}
}
\end{table}
\end{widetext}

These $192$ equations can be condensed into one compact equation, namely:

\begin{equation}
iA^{\mu}\gamma^{\mu}_{(a)}\partial_{\mu}\psi=(-1)^{\ell+l}(i)^{l}\left(\frac{m_{0}c}{\hbar}\right)\tilde{\gamma}^{\ell}\psi_{(j)},\label{aldirac}
\end{equation}

where $l=0,1$; $j=1,2$; $\ell=0,1,2,3, ..., 15$; $a=1,2,3$. For $j=1$, we have $\psi_{(1)}=\psi$; and for $j=2$, we have $\psi_{(2)}=\psi_{c}$; and for $a=1$ we have $\gamma^{\mu}_{(1)}=\gamma^{\mu}$; for $a=2$ we have $\gamma^{\mu}_{(2)}=\bar{\gamma}^{\mu}$, and finally, for $a=3$ we have $\gamma^{\mu}_{(3)}=\hat{\gamma}^{\mu}$. In table (\ref{equations}), by the ``Nature of Spacetime'' the refers to whether $\lambda=\pm1,0$ and as will be since in \S (\ref{invrestmass}) $\lambda=0$ for the QST, $\lambda=+1$ for the HST, and $\lambda=-1$ for the PST  and; the superscript $(+1)$, $(-1)$ and $(0)$ in $E^{(+1)}_{+}$,$E^{(+1)}_{-}$, $E^{(-1)}_{+}$, $E^{(-1)}_{-}$,$E^{(0)}_{+}$ and $E^{(0)}_{-}$ indicates that this is the energy solution for the HST, PST and QST respectively, and subscript $\pm$  indicates that this is the positive or negative energy solution respectively.

If one where to follow the same procedure as in Nyambuya ($2007$), they should be able to show that the formula for the anomalous gyromagnetic ratio emerging from the the above Curved Spacetime Dirac Equations is the same  as that derived in Nyambuya ($2007$), \textit{i.e.}, $\Delta a^{(\pm1)}$ and $\Delta a^{(0)}$ for the QST-Eqns and DCST-Eqns respectively.

Further, we did show in Nyambuya ($2009$) that one can write the Dirac Equation such that it describes in general any spin particle. This was achieved by modifying the $\gamma^{\mu}_{(a)}$-matrices. As equation (\ref{aldirac}) stands, it only describes spin-$1/2$ particles. If we want it to describe any spin particle, we will have to modify (as we did to the Dirac Equation in Nyambuya $2009$) the matrices $\gamma^{\mu}_{(a)}$; from $\gamma^{\mu}_{(a)}$ to $\gamma^{\mu}_{(as)}$ where the new label $s=0,1,2,3,...$ labels the spin, such that:

\begin{equation}
\begin{array}{c c}
\gamma^{0}_{(1s)}=
\left(\begin{array}{c c}
\textbf{I} & \mathbf{0}\\
\mathbf{0} & -\textbf{I} \\
\end{array}\right)
,\,\,\,\,
\gamma^{i}_{(1s)}=
s\left(\begin{array}{c c}
\textbf{0} & \boldsymbol{\sigma}^{i}\\
-\boldsymbol{\sigma}^{i} & \textbf{0} \\
\end{array}\right),
\end{array}\label{gamma-ms}
\end{equation}

\begin{equation}
\begin{array}{c c}
\gamma^{0}_{(2s)}=
\left(\begin{array}{c c}
\textbf{I} & \mathbf{0}\\
\mathbf{0} & -\textbf{I} \\
\end{array}\right)
,\,\,\,\,
\gamma^{i}_{(2s)}=
\frac{1}{2}s\left(\begin{array}{c c}
2\textbf{I} & i\sqrt{2}\boldsymbol{\sigma}^{i}\\
-i\sqrt{2}\boldsymbol{\sigma}^{i} & -2\textbf{I} \\
\end{array}\right),
\end{array}\label{gamma-bar-ms}
\end{equation}

\begin{equation}
\begin{array}{c c}
\gamma^{0}_{(3s)}=
\pm\left(\begin{array}{c c}
\textbf{I} & \mathbf{0}\\
\mathbf{0} & -\textbf{I} \\
\end{array}\right)
,\,\,\,\,
\gamma^{i}_{(3s)}=
\mp\frac{1}{2}s\left(\begin{array}{c c}
2\textbf{I} & i\sqrt{2}\boldsymbol{\sigma}^{i}\\
-i\sqrt{2}\boldsymbol{\sigma}^{i} & -2\textbf{I} \\
\end{array}\right),
\end{array}\label{gamma-hat-ms}
\end{equation}

hence:

\begin{equation}
iA^{\mu}\gamma^{\mu}_{(as)}\partial_{\mu}\psi=(-1)^{\ell+l}(i)^{l}\left(\frac{m_{0}c}{\hbar}\right)\tilde{\gamma}^{\ell}\psi_{(j)},\label{aldiracs}
\end{equation}

is a General Spin Curved Spacetime Dirac Equation and thus far, this is the most complete Curved Spacetime Equation that we have. In Nyambuya (2009), we shall make another modification to the Curved Spacetime Dirac Equation, \textit{i.e.}, we shall include a four vector cosmological field.

\section{\sectionfont Some Properties of the Curved Spacetime Dirac Equations\label{props}}

\subsection{\subsectionfont Invariance Under Lorentz Transformations\label{linv}}

Proving the Lorentz invariance of just one of the twelve equations, is as good as proving the Lorentz invariance of the rest of the equations as this same procedure is what must be used for proving the Lorentz invariance of the rest of the equations. For clarity reasons, we shall drop the $as$-indices in (\ref{aldiracs}). Now, taking equation (\ref{pdirac1}) multiplying it by $\hbar\tilde{\gamma}^{\ell} $ and after rearranging,  we have:

\begin{equation}
\left[\left(-1\right)^{\ell}i\hbar A^{\mu}\tilde{\gamma}^{\ell}\bar{\gamma}^{\mu}\partial_{\mu} -m_{0}c\right]\psi=0.
\end{equation}

To avoid confusing the term $A^{\mu}\tilde{\gamma}^{\ell}\bar{\gamma}^{\mu}$ as a product of three vectors $A^{\mu}$, $\tilde{\gamma}^{\ell}$ and $\bar{\gamma}^{\mu}$, it is best to write this as $\Gamma^{\ell\mu}=A^{\mu}\tilde{\gamma}^{\ell}\bar{\gamma}^{\mu}$. Only $A^{\mu}$ is the vector and $\tilde{\gamma}^{\ell}$ and $\bar{\gamma}^{\mu}$ are constant matrices that are frame-independent ($\bar{\gamma}^{\mu}$ are pure constants and as will be demonstrated and for the case $\tilde{\gamma}^{\ell}$, these matrices are by definition pure constants and independent of the frame of reference) hence $\Gamma^{\ell\mu}$ are 16, $4\times4$ four-vector matrices thus the above  equation can written as:

\begin{equation}
\left[\left(-1\right)^{\ell}i\hbar \Gamma^{\ell\mu}\partial_{\mu} -m_{0}c\right]\psi=0.
\end{equation}

Written in this manner and given that $\Gamma^{\ell\mu}$ are sixteen $4\times4$ four-vector matrices -- \textit{to the agile}; the Lorentz invariance of this equation is clearly visible. But for the sake of formalities, we shall proceed to show its Lorentz invariance and the reader must not forget that $A^{\mu}\tilde{\gamma}^{\ell}\bar{\gamma}^{\mu}$ are $4\times4$ four-vector matrices, otherwise one runs into trouble. In our derivation of the Lorentz invariance, we shall not write this $4\times4$ four-vector matrices as $\Gamma^{\ell\mu}$ in case the matrices $\bar{\gamma}^{\mu}$ are frame dependent -- we  show that they are not.

To prove Lorentz covariance, two conditions must be satisfied, \textit{i.e.}:

\textbf{1.} Given any two inertial observers $\textrm{O}$ and $\textrm{O}^\prime$ anywhere in spacetime, if in the frame $\textrm{O}$ we have $[\left(-1\right)^{\ell}i\hbar A^{\mu}\tilde{\gamma}^{\ell}\bar{\gamma}^{\mu}\partial_{\mu}-m_{0}c]\psi(x)~=~0$, then $[\left(-1\right)^{\ell}i\hbar A^{\prime\mu}\tilde{\gamma}^{\prime \ell}\bar{\gamma}^{\prime\mu}\partial_{\mu}^\prime~-~m_{0}c]\psi^\prime(x^\prime)~=~0$ is the equation describing the same state but in the frame $\textrm{O}^\prime$. 

\textbf{2.} Given that $\psi(x)$ is the wavefunction as measured by observer $\textrm{O}$, there must be a prescription for observer $\textrm{O}^\prime$ to compute $\psi^\prime(x^\prime)$ from $\psi(x)$ and this describes to $\textrm{O}^\prime$ the same physical state as that measured by $\textrm{O}$.

Now, since the Lorentz transformations are linear transformations, it is to be required or expected of the transformations between $\psi(x)$ and $\psi^\prime(x^\prime)$  to be linear too, \textit{i.e.}:

\begin{equation} 
\psi^\prime(x^\prime) = \psi^\prime(\Lambda x) = S(\Lambda) \psi(x) = S(\Lambda) \psi(\Lambda^{-1}x^\prime),\label{inverse1}
\end{equation} 

where $S(\Lambda)$ is a $4\times 4$ matrix which depends only on the relative velocities of $\textrm{O}$ and $\textrm{O}^\prime$. $S(\Lambda)$ has an inverse if $\textrm{O}\rightarrow \textrm{O}^\prime$ and also $\textrm{O}^\prime\rightarrow \textrm{O}$. The inverse is:

\begin{equation} 
\psi(x) = S^{-1}(\Lambda)\psi^\prime(x^\prime) = S^{-1}(\Lambda)\psi^\prime(\Lambda x), \label{inverse2}
\end{equation} 	

or we could write:

\begin{equation}
\psi(x)=S(\Lambda^{-1})\psi^\prime(\Lambda x)\Longrightarrow S(\Lambda^{-1}) = S^{-1}(\Lambda),
\end{equation}

We can now write $[\left(-1\right)^{\ell}i\hbar A^{\mu}\tilde{\gamma}^{\ell}\bar{\gamma}^{\mu}\partial_{\mu}-m_{0}c]\psi(x)=0$ 
as  $[\left(-1\right)^{\ell}i\hbar A^{\mu}\tilde{\gamma}^{\ell}\bar{\gamma}^{\mu}\partial_{\mu}-m_{0}c]S^{-1}(\Lambda)\psi^\prime(x^\prime)=0$ and multiplying this from the left by $S(\Lambda)$ we have $S(\Lambda)[\left(-1\right)^{\ell}i\hbar A^{\mu}\tilde{\gamma}^{\ell}\bar{\gamma}^{\mu}\partial_{\mu}-m_{0}c] S^{-1}(\Lambda)\psi^\prime(x^{\prime})=0$
and hence:

\begin{equation}
\left[\left(-1\right)^{\ell}i\hbar S(\Lambda)\tilde{\gamma}^{\ell}\bar{\gamma}^{\mu} S^{-1}(\Lambda)A^{\mu}\partial_\mu - m_{0}c\right]\psi^\prime(x^\prime)=0.
\end{equation}

Now, since $A^{\mu}$ is a vector, it is clear that $A^{\mu}\partial_\mu$ is a scalar, \textit{i.e.}, $A^{\mu}\partial_{\mu}=A^{\prime\mu}\partial^{\prime}_{\mu}$, therefore we will have:

\begin{equation}
\left[\left(-1\right)^{\ell}i\hbar S(\Lambda)\tilde{\gamma}^{\ell}\bar{\gamma}^{\mu} S^{-1}(\Lambda)A^{\prime\mu}\partial^{\prime}_{\mu} - m_{0}c\right]\psi^\prime(x^\prime)=0.
\end{equation}

Therefore the requirement is that $\tilde{\gamma}^{\prime \ell}\bar{\gamma}^{\prime\mu}=S(\Lambda)\tilde{\gamma}^{\ell}\bar{\gamma}^{\mu} S^{-1}(\Lambda)$ and since $\tilde{\gamma}^{\prime \ell}=\tilde{\gamma}^{\ell}$ (because the matrices $\tilde{\gamma}^{\ell}$ are pure constants) and $\left(\tilde{\gamma}^{\ell}\right)^{-1}=\tilde{\gamma}^{\ell\dagger}$, from this we will have: $\bar{\gamma}^{\prime\mu}=\tilde{\gamma}^{\ell\dagger}S(\Lambda)\tilde{\gamma}^{\ell}\bar{\gamma}^{\mu} S^{-1}(\Lambda)$ which further reduces to: $\bar{\gamma}^{\prime\mu}=S^{\dagger}(\Lambda)\tilde{\gamma}^{\ell}\tilde{\gamma}^{\ell}\bar{\gamma}^{\mu} S^{-1}(\Lambda)$ and given that: 
$\tilde{\gamma}^{\ell}\tilde{\gamma}^{\ell}=\textrm{I}$ this means: $\bar{\gamma}^{\prime\mu}=S^{\dagger}(\Lambda)\bar{\gamma}^{\mu} S^{-1}(\Lambda)$ and lastly, this further reduces to: $\bar{\gamma}^{\prime\mu}=S^{\dagger}S^{-1\dagger}(\Lambda)(\Lambda)\bar{\gamma}^{\mu\dagger} $ and since $\bar{\gamma}^{\mu\dagger}=\bar{\gamma}^{\mu}$, our final result from all these computations is: $\bar{\gamma}^{\prime\mu}=\bar{\gamma}^{\mu}$. This result can be generalized to:

\begin{equation} 
\gamma^{\prime\mu}_{(as)}=\gamma^{\mu}_{(as)}. 
\end{equation}

This is a very important result as it tells us that the matrices $\gamma^{\mu}_{(as)}$ are pure constants, there are -- unlike in the Dirac Theory; independent of the frame of reference. If the  requirement is made that  $S(\Lambda)$ form a representation of the Lorentz group, this relationship defines $S(\Lambda)$ only up to an arbitrary factor and this factor is restricted to a $\pm$ sign. With this, we obtain the two-valued spinor representation and wave-function transforming according to equation (\ref{inverse1}). With this we have shown the Lorentz invariance of equation (\ref{pdirac1}) hence thus as initially argued, we have shown that all the rest of equations are Lorentz invariance as this same method applies to the rest of the equations in proving the Lorentz invariance  these equations. 

\subsection{\subsectionfont Invariance Under Rest Mass Reversal\label{invrestmass}}

By studying the properties of the energy equation under different operations, we can deduce the properties of the Curved Spacetime Dirac Equation. Having deduced these, the next thing is simple to verify them rigorously. Equations (\ref{diracf1}), (\ref{diracf2}), (\ref{ndiracc1}), (\ref{ndiracc2}), (\ref{ndirac1}), (\ref{ndirac2}), (\ref{nndirac1}), (\ref{nndirac2}), (\ref{pdirac1}), (\ref{pdirac2}), (\ref{npdirac1}), and (\ref{npdirac2}) satisfy the energy equation for Curved Spacetime is given by:

\begin{equation}
(A^{0})^{2}E^{2}-\left(2\lambda A^{0}A^{k} p_{k}c\right)E-A^{j}A^{k}p_{j}p_{k}c^{2} = m_{0}^{2}c^{4},\label{energy1}
\end{equation}

and just to clean-up this messy equation lets set $\mathcal{M}_{0}=m_{0}/A^{0}$, $\mathcal{P}=A^{k}p_{k}/A^{0}$  ($k=1,2,3$) and realizing that $A^{j}A^{k}p_{j}p_{k}= (A^{k}p_{k})^{2}=(A^{0})^{2}\mathcal{P}^{2}$, the solution to equation (\ref{energy1}) with respect to $E$ is given by:

\begin{equation}
E=\lambda \mathcal{P}c\pm \sqrt{\left(1+\lambda^{2}\right)\mathcal{P}^{2}c^{2}+\mathcal{M}^{2}_{0}c^{4}}\label{energy2},
\end{equation}

where $\lambda=\pm1,0$ and the case $\lambda=0$ is for the QST, and $\lambda=+1$ is the case for the HST and likewise $\lambda=-1$ is the case for the PST. This energy equation, \textit{i.e.} (\ref{energy2}), is the energy equation for the case of only spin-1/2 particles \textit{i.e.}, those described by (\ref{aldirac}). For the general spin case, \textit{i.e.}, for particles described by (\ref{aldiracs}), the energy is given by:

\begin{equation}
E(s)=\lambda s\mathcal{P}c\pm \sqrt{\left(1+\lambda^{2}\right)s^{2}\mathcal{P}^{2}c^{2}+\mathcal{M}^{2}_{0}c^{4}}\label{energy3},
\end{equation}

Now, from equation (\ref{energy1}), it is clear that $m_{0}\longmapsto-m_{0}$ leaves this equation uncharged. This means under the interchange of the rest-mass, the QST-Eqns, HST-Eqns and PST-Eqns must remain invariant as-well. We shall investigate this in the subsequent subsections.

Now, considering (\ref{energy3}), for the case $s=1$ and $\mathcal{P}>0$, and considering only the positive energy solutions, \textit{i.e.}, $E>0$, we will have for $\lambda=0$, $E^{(0)}_{+}=\sqrt{\mathcal{P}^{2}c^{2}+\mathcal{M}^{2}_{0}c^{4}}$, and for $\lambda=-1$, $E^{(-1)}_{+}=-\mathcal{P}c+\sqrt{2\mathcal{P}^{2}c^{2}+\mathcal{M}^{2}_{0}c^{4}}$, and for $\lambda=+1$, $E^{(+1)}_{+}=+\mathcal{P}c+\sqrt{2\mathcal{P}^{2}c^{2}+\mathcal{M}^{2}_{0}c^{4}}$. From this clearly, we will have:

\begin{equation}
E^{(0)}_{+}<E^{(-1)}_{+}<E^{(+1)}_{+}\Longrightarrow E^{(0)}_{-}>E^{(-1)}_{-}>E^{(+1)}_{-}\label{ehierarchy},
\end{equation}

Clearly this points to a family of particles with a three stage hierarchy noted by their energies and hence mass. This is what we shall suggest in \S (\ref{lgp}) as the reason the reason behind the ``mysterious'' lepton generation problem. We will leave the case for quarks, for a latter reading, but it should be said here that, certainly, this same mechanism is what gives rise to the particle generation phenomena.
\subsubsection{\subsubsectionfont Case I}

After the transformation $m_{0}\longmapsto-m_{0}$, for all the equations, \textit{i.e.,} the QST-Eqns and the QST-Eqns, we can revert back to the original equation  by a simultaneous reversal of the space and time coordinates, \textit{i.e.}, $t\longmapsto -t$ and $x^{k}\longmapsto -x^{k}$ ($k=1,2,3$). If as before, $\left<E_{+}, E_{-}\right>$ is the ordered pair of the energy solutions for these equations with $E_{+}>0$ and $E_{-}<0$, the transformation $t\longmapsto -t$ and $x^{k}\longmapsto -x^{k}$ flips the energy to a different order, \textit{i.e.}, $\left<|E_{-}|, -E_{+}\right>$. In a nutshell, this means, the equation is not invariant under these transformations. 

For the QST-Eqns, one can revert back to the original equation by taking the complex conjugate on both sides and then multiplying throughout by $\gamma^{0}\gamma^{2}$ and then re-arranging the matrices. This set of operations will be demonstrated in $\S$ (\ref{cconj}). \textit{Sadly}, the same set of operations does not leave the DCST-Eqns invariant. From the vantage-point of beauty, simplicity and (computational) economy, it would be unacceptable for these equations to be invariant under the reversal of the rest-mass through a mirrad of operations if one can find just one such operation which does the same job for all the the equations in one full-swap. By mirrad of operations, it is meant that there is a set of operations that applies to the QST-Eqns leaving them invariant and another set that applies to the DCST-Eqns for the same job. If one operation can do the job for both the QST-Eqns and the DCST-Eqns, this set of operations is the superior of them and must chosen -- computationally, it is the ``computational geodesic path''. It so happens, that one such operation exists. This, we will show next.

\subsubsection{\subsubsectionfont Case II}

After the transformation $m_{0}\longmapsto-m_{0}$, for all the equations, \textit{i.e.,} the QST-Eqns and the DCST-Eqns, we can revert back to the original equation  by reversing the Electromagnetic field, \textit{i.e.,} $A^{\mu}\longmapsto -A^{\mu}$ and unlike in Case I, this does not flip the energy of the particles, hence this is the transformation we are seeking. The fact that $m_{0}\longmapsto -m_{0}\Longrightarrow A^{\mu}\longmapsto -A^{\mu}$ clearly points to the existence of an intimate relationship between the rest-mass of a particle and its electronic charge, otherwise how can one explain the automatic flipping of the sign of the rest-mass when the Electromagnetic field is reversed? At the very least, this relationship must be a direct proportionality relationship, or an odd power direct proportionality relationship, \textit{i.e.}:

\begin{equation}
m_{0}\propto Q^{2n+1};\,n=0,1,2,3,4, \textbf{...}\,etc, \label{mq}
\end{equation}

where $Q$ is the electronic charge of the particle in question. For this kind of setting, since the rest-mass has an odd-power direct proportionality relationship to the electronic charge, a change in the sign of the Electromagnetic field, will automatically lead to a reversal of the sign of the rest-mass if the Electromagnetic field is reversed. For simplicity, we shall assume from here-on that $m_{0}\propto Q$. In \S (V C) of the reading Nyambuya (2007; latest version) where the Curved Spacetime Dirac Equation is derived from the soils of UFT, the view that $m_{0}\propto Q$ is justified.

If this is correct, the meaning \textit{vis} matter/antimatter relationship, is that the antiparticle of a positive energy-mass particle has positive energy-mass as-well and not negative energy has the Dirac Theory implies. Thus, once again, if this analysis is correct (and off cause the theory as-well), the question of whether antiparticles would in (say) the gravitational field of the earth fall-up instead of down, may have found an answer. 

It is important to note that, if the above is correct, then,  the rest-mass $m_{0}$ and the mass $m=E/c^{2}$ will have a different meanings from that currently understood. This issue will be tackled in a separate future reading.

\subsection{\subsectionfont Invariance Under Charge Conjugation\label{cconj}}

Invariance under charge conjugation is a symmetry ushered into physics by Paul A. M. Dirac \textit{via} his Dirac Equation. This invariance -- contrary to observations, entails that the Universe ought to be filled with equal portions of matter and antimatter. We have argued in the reading Nyambuya ($2009$; which is yet to be given a final conclusion) that the inclusion of a four vector cosmological field into the spacetime dimensions potentially explains why the Universe appears to be asymmetric in its matter-antimatter distribution. With regard to electronic charge conjugation symmetry, we shall require of all our equations, \textit{i.e.}, equation (\ref{aldirac}) to behave like the original Dirac Equation thus obey the electronic charge conjugation symmetry. As has been done to the Dirac Equation in the reading Nyambuya ($2009$), we will expect the electronic charge conjugation symmetry to be broken only upon the inclusion of the four vector cosmological field, thereby potentially explaining  why the Universe appears to dominated by matter and not antimatter.

In the Dirac sense -- \textit{viz}, the relationship between particle and antiparticle, for each particle, there exists an antiparticle where the antiparticle has the same properties as the particle except that its electronic charge, mass and energy are the exact opposite to that of the particle. However, the idea that antiparticles have a negative mass and energy is not settled and is treated with great care. No experiment to date has ever tested this perdurable feature of the Dirac Theory -- if perhaps by our own shear ignorance such an experiment has been conducted, we are sure it has not delivered a vindicative answer because whether the result is positive or negative, this would make great news for the physics community. 

In this same vein of quest, of whether or not antiparticles have a negative answer, in the Dirac theory -- \textit{e.g.}, the existence of the electron ($\textrm{e}^{-}$) implies the existence of the positron ($\textrm{e}^{+}$) whose energy is negative and in the present as implied by the Curved Spacetime Dirac Equation, we shall show that invariance under charge conjugation holds only if we reset the rest-mass  of the particle, thus strongly suggesting that the rest-mass of a particle ought to have an intimate and direct proportionality relationship with electronic charge ($m_{0}\propto Q$). If this is the case, we are brought closer to answering the question of whether antiparticles have a negative mass and energy. We shall also see that this finding that the rest-mass of a particle may well be related to the electronic charge of the particle leads us to a possible answer as to whether or not neutrinos have a rest-mass! 

To show this -- \textit{i.e.}, the invariance under charge conjugation -- we proceed as usual taking equation (\ref{pdirac1}) is an example. First we bring the particle under the influence of an external Electromagnetic magnetic field $A_{\mu}^{ex}$ (which is a real function); having done this, the normal procedure  is to make the transformation $\partial_{\mu} \longmapsto \textrm{D}_{\mu}=\partial_{\mu}+iA_{\mu}^{ex}$ hence equation (\ref{pdirac1}) will now be given by

\begin{equation}
\left[i\hbar A^{\mu}\bar{\gamma}^{\mu}\textrm{D}_{\mu}- \left(-1\right)^{\ell}\tilde{\gamma}^{\ell}m_{0}c\right]\psi=0.\label{ndirac}
\end{equation}

Now, we shall for our own convenience rewrite equation (\ref{ndirac})  in the form:

\begin{equation}
A^{0}\gamma^{0}\textrm{D}_{0}\psi-i\left(\frac{\sqrt{2}}{2}\right)A^{k}\gamma^{k}\textrm{D}_{k}\psi +\gamma^{0}A^{k}\textrm{D}_{k}\psi=\left(-1\right)^{\ell}\left(\frac{m_{0}c}{\hbar}\right)\tilde{\gamma}^{\ell}\psi.\label{mdirac3}
\end{equation}

In this form we have written (\ref{ndirac}) in-terms of the usual $\gamma^{\mu}$-matrices  because these $\gamma^{\mu}$-matrices are easier to manipulate as we are more used to them than the $\bar{\gamma}^{\mu}$ and $\hat{\gamma}^{\mu}$-matrices. 

Proceeding, we take the complex conjugate on both-sides of this equation, we will have:

\begin{equation}
A^{0}\gamma^{0}\textrm{D}_{0}^{*}\psi^{*}+i\left(\frac{\sqrt{2}}{2}\right)A^{k}\gamma^{k*}\textrm{D}_{k}^{*}\psi^{*} +\gamma^{0}A^{k}\textrm{D}_{k}^{*}\psi^{*}=\left(-1\right)^{\ell}\left(\frac{m_{0}c}{\hbar}\right)\tilde{\gamma}^{\ell*}\psi^{*},
\end{equation}

and then multiply this by $\gamma^{0}\gamma^{2}$ and then using the relations:

\begin{equation}
\begin{array}{l r}
\left\{\gamma^{\mu},\gamma^{2}\right\}=0 ;\,\mu\neq2 & \dots \,(\textbf{a})\\
\\
\left\{\gamma^{0},\gamma^{k}\right\}=0:\,k=1,2,3 & \dots \,(\textbf{b})\\
\end{array}\label{mrel}
\end{equation}

where $\left\{\right\}$ is the anti-commutator bracket, we are lead to:

\begin{equation}
\left[i\hbar A^{\mu}\bar{\gamma}^{\mu}\textrm{D}_{\mu}^{*}- \left(-1\right)^{\ell}\tilde{\gamma}^{\ell*}m_{0}c\right]\psi_{c}=0.
\end{equation}

Now, $\textrm{D}^{*}_{\mu}=\partial_{\mu}-iA_{\mu}^{ex}$; if we reverse  the external  Electromagnetic field, $A_{\mu}^{ex}\longmapsto-A_{\mu}^{ex}$, we also have to reverse that of the particle, \textit{i.e.}, $A^{\mu}\longmapsto-A^{\mu}$ and we must remember (equation \ref{mq}) that a reversal of the Electromagnetic field is intimately coupled to a reversal of the rest-mass hence $m_{0}\longmapsto-m_{0}$  and  since $\tilde{\gamma}^{\ell}$ is real, this means $\tilde{\gamma}^{\ell*}=\tilde{\gamma}^{\ell}$ and also we see that:

\begin{equation}
\begin{array}{l r}
\left\{\gamma^{0}\gamma^{2},\tilde{\gamma}^{\ell}\right\}=0: \ell=2n+1 & \dots \,(\textbf{a})\\
\\
\left[\gamma^{0}\gamma^{2},\tilde{\gamma}^{\ell}\right]=0:\,\ell=2n & \dots \,(\textbf{b})\\
\end{array}
\end{equation}

and effecting all these, we will have:

\begin{equation}
\left[i\hbar A^{\mu}\bar{\gamma}^{\mu}\textrm{D}_{\mu}- \left(-1\right)^{\ell}\tilde{\gamma}^{\ell}m_{0}c\right]\psi_{c}=0\label{cdirac2},
\end{equation}

which is the original equation and this completes the proof that equation (\ref{pdirac1}) is invariant under charge conjugation. Performing the same operations to all the other DCST-Eqns leads to the same conclusion hence thus all the DCST-Eqns obey charge conjugation symmetry as-well.

For the case of the QST-Eqns under the influence of an ambient Electromagnetic field to prove the invariance under charge conjugation, we;  (1) make the necessary transformation $\partial_{\mu}\longmapsto \partial_{\mu}+iA_{\mu}^{ex}$, (2) take the the complex conjugate on both sides of the equations, (3)  multiply both-sides by $\gamma^{0}\gamma^{2}$, (4) make the necessary algebra using equation (\ref{mrel}) (a) to rearrange the matrices and restore the $\pm$ signs to their  original settings on both the left and right hand side, (5) reverse  the external  Electromagnetic field, $A_{\mu}^{ex}\longmapsto-A_{\mu}^{ex}$ and that of the particle, $A^{\mu}\longmapsto-A^{\mu}$ and at the sametime remembering (equation \ref{mq}) that a reversal of the Electromagnetic field comes along with the reversal of the rest-mass $m_{0}\longmapsto-m_{0}$,  and  also noting that $\tilde{\gamma}^{\ell}$ is real meaning $\tilde{\gamma}^{\ell*}=\tilde{\gamma}^{\ell}$  and (6) having gone through 1 to 5 correctly, one must have the original QST-Eqns, hence we will have shown or proved the invariance of the QST-Eqns under charge conjugation.

\subsection{\subsectionfont Symmetry Under  Space and Time Inversion}

We proceed to investigate another of the symmetries -- invariance  under space (otherwise also known as parity and symbolized the letter $\textrm{P}$) and time $\textrm{T}$ inversion or the lack thereof.  Starting with space inversion, simple, space inversion is the transformation of the space coordinates $x^{i}\longmapsto -x^{i}$ ($i=1,2,3$) and this implies $\partial_{i}\longmapsto -\partial_{i}$ and making this transformation into the  QST-Eqns, we can revert back to the original equations by first taking the complex conjugate on both sides of these equations before multiplying by $\gamma^{0}\gamma^{2}$  and making the necessary algebra as is done above in \S (\ref{cconj}). Hence thus, the flat spacetime equations are invariant under space inversion. In the case of the DCST-Eqns, it is not possible to revert back to the original equation as is the case for the flat spacetime equation above. In a nutshell, the DCST-Eqns are not invariant under space inversion.

Proceeding to the translations under time interchange, \textit{i.e.,} $t\longmapsto-t$, it is seen that the QST-Eqns are invariant under time translations and as before the DCST-Eqns are not invariant under time reversal. The same goes for simultaneous translation of both space and time, \textit{i.e.,} $x^{\mu}\longmapsto -x^{\mu}$, the  QST-Eqns are invariant while the  DCST-Eqns are not invariant.

It is relatively easy and straight forward to show that the combined charge, space and time reversal symmetries -- namely $\textrm{CPT}$; is not violated by all the equations. After making the necessary transformations, one simple has to take the complex conjugate on both sides of these equations before multiplying by $\gamma^{0}\gamma^{2}$ and this is for the QST-Eqns and in the case of the DCST-Eqns one simple has after taking the complex conjugate on both sides of these equations, to multiply by $\gamma^{0}\gamma^{2}$  and making the necessary algebra as in \S (\ref{cconj}), in order to revert back to the original equation.

\section{\sectionfont CP Violation\label{cpv}}

$\textrm{CP}$-\textit{symmetry}, the product of the two discrete symmetries $\textrm{C}$ and $\textrm{P}$. This symmetry was thought to restore order after $\textrm{P}$-symmetry violation was discovered in the now famous $1956$ experiments proposed by Tsung-Dao Lee and Chen Ning Yang where carried by a group led by Chien-Shiung Wu. The Strong  and Electromagnetic interaction seem to be invariant under $\textrm{CP}$ transformation operation, but this symmetry is violated by certain types of weak interactions. 

Using the same procedures as above, it is not difficult to see that the DCST-Eqns  will all violate $\textrm{CP}$-\textit{symmetry}. All known and accepted equations in physics that describe particles (\textit{e.g.} Dirac Equation, Schr\"odinger Equation, Proca Equation, Klein Gordon) in their bare and natural form do not violate this symmetry and in order for there to be $\textrm{CP}$-\textit{symmetry} these equations must be modified (in the case of the weak interaction under the bare and natural Dirac Equation modification to this equation are needed) to fit observations of this $\textrm{CP}$-symmetry violation.  With the DCST-Eqns, this is wholly part and parcel of the natural fabric of these equations hence, if these equation correspond to natural reality (as I would like to believe) a natural explanation for  $\textrm{CP}$-\textit{symmetry} violation, may have for the first time found a natural home as a consequence of the curvature of spacetime.ca Equation, Klein Gordon) in their bare and natural form do not violate this symmetry and in order for there to be $\textrm{CP}$-\textit{symmetry} these equations must be modified (in the case of the weak interaction under the bare and natural Dirac Equation modification to this equation are needed) to fit observations of this $\textrm{CP}$-symmetry violation.  With the DCST-Eqns, this is wholly part and parcel of the natural fabric of these equations hence, if these equation correspond to natural reality (as I would like to believe and this belief stems from the equations' simplicity and beauty) a natural explanation for  $\textrm{CP}$-\textit{symmetry} violation, may have for the first time found a natural home as a consequence of the nature of the curvature of spacetime \textit{i.e.}, is the spacetime a QST, HST or PST? If the spacetime is a HST or PST, then $\textrm{CP}$--violation will occur.

$\textrm{CP}$-\textit{symmetry} if it where a symmetry of nature, implies that the equations of particle physics are invariant under mirror inversion and this leads naturally to the prediction that the mirror image of a reaction (such as a chemical reaction or radioactive decay) should occur at the same rate as the original reaction. Not until $1956$,  along with conservation of energy and conservation of momentum, $\textrm{CP}$-\textit{symmetry}, was believed to be one of the fundamental geometric Conservation Laws \textit{of} Nature. As has already been mentioned, this changed after a careful critical review of the existing experimental data by Tsung-Dao Lee and Chen Ning Yang in $1956$.

Tsung-Dao Lee and Chen Ning Yang, after realizing that while experiments had revealed  that $\textrm{CP}$-symmetry had been verified in decays by the Strong or Electromagnetic interactions, it was untested in the Weak interaction, proposed several possible direct experimental tests on the Weak interactions, the first being on beta decay of Cobalt-60 nuclei and this was carried out in $1956$ by a group led by Chien-Shiung Wu, and this demonstrated conclusively that weak interactions indeed violate the $\textrm{P}$-\textit{symmetry}y. This was inferred from the analogy that some reaction of the Weak interactions did not occur as often as their mirror image did as would be expected if $\textrm{P}$-symmetry where conserved.

Directly connected with $\textrm{CP}$ violation, is the major unsolved problem in theoretical of why the universe seems to be made-up chiefly of matter, rather than consisting of equal parts of matter and antimatter. It can be demonstrated, as was done by Sakharov ($1967$), that to create an imbalance in matter and antimatter from an initial condition of balance, certain conditions must be meet and these conditions have come to the called the Sakharov conditions and $\textrm{CP}$-violation is one of the conditions.
 
The Big Bang, which (at present) is believed to have brought forth the Universe into being, should -- according the our current understanding of the symmetries seen to be exhibited by the Laws \textit{of} Nature; have produced equal amounts of matter and anti-matter if $\textrm{CP}$-\textit{symmetry} was preserved hence thus, there should have been total cancellation of both. In other words, protons should have canceled with anti-protons, electrons with positron, neutrons with anti-neutrons, and so on for all elementary particles. This would have resulted in a sea of photons in the Universe -- this means a Universe devoid of any form of matter. Since this is quite evidently not the case, after the Big Bang, Physical Laws must have acted differently for matter and antimatter, \textit{i.e.}, violating $\textrm{CP}$-\textit{symmetry} -- \textit{so it is thought}. How this $\textrm{CP}$-symmetry violation would come about is not exactly known. 

From a different point of departure, in \S (V C) of the reading Nyambuya (2007; latest version) where the Curved Spacetime Dirac Equation is derived from the soils of UFT, this question of the dominance of matter over antimatter is tackled. We will not attempt to answer this question. We will do so in Nyambuya ($2009a$). There in Nyambuya ($2009a$), it will be shown that $\textrm{CP}$-\textit{symmetry} violation has little -- if at all anything; to do with  this mystery of matter-antimatter imbalance.

\section{\sectionfont Lepton Generation Problem\label{lgp}}

According to current wisdom, leptons have three generations and these generations are notably  marked by their masses. Each generation is divided into two leptons and the two leptons may be divided into one with electric charge $-1$ and one electrically neutral particle -- the neutrino. As shown in  table (\ref{lep}), the first generation consists of the Electron, Electron-neutrino, \textit{i.e.}, ($\textrm{e}^{-},\nu_{e}$). The second generation consists of the Muon, Muon-neutrino, \textit{i.e.}, ($\mu^{-},\nu_{\mu}$). The third generation consists of the Tau lepton, Tau-neutrino, \textit{i.e.}, ($\tau^{-},\nu_{\tau}$). Each member of a higher generation has greater mass than the corresponding particle of the previous generation. The question as to why this is so -- \textit{in my view}; is perhaps well summarized by the words of Veltman ($2003$):

\begin{quote}
\textit{``Perhaps the greatest mystery of them all is the remarkable three-family structure of quarks and leptons. No one has found any explanation for this structure. We are reasonable sure, that there are no more than three families.''}
\end{quote} 

As to the last words in the above words of Veltman, \textit{i.e.}; ``\textit{We are reasonable sure, that there are no more than three families.}'', we will have to wait until the reading Nyambuya ($2009a$) is complete before we can say our opinion on that matter as it appears the ideas in the  reading Nyambuya ($2009a$) seem to point at a possible existence of other Leptons -- so, we must wait until this reading is brought to a halt. 

To the other words of Veltman, \textit{i.e.}; ``\textit{No one has found any explanation for this structure ...}''; a suggestion is here made that tries to explain this three-family structure using the Curved Spacetime Dirac Equations.  By no means do we say this is the explanation, but this is simple a hint at the solution. Because of the findings in the reading Nyambuya (2009) which is still to be given a final conclusion, we are aware that this reading suggests the existence of there more leptons, hence in the light of this, it is pretty much premature to make any conclusions in the present. 

From the energy inequality (\ref{ehierarchy}), we have shown that there will exist a three stage energy hierarchy -- and from this; if the mass of particles is given by the energy equivalent $m=E/c^{2}$ as in the Einstein sense of mass-energy equivalence, then, there must exist a family of particle with three stage mass structure; this points to a three-member hierarchy of particles in terms of mass and it is this observations that we will seize upon and use to suggest a solution to the generation problem of fermions.

\begin{table}[!h]
\caption{\tabletitlefont Leptons}\label{lep}
\vspace*{0.3cm}
\tabletextfont
\begin{tabular}{ c l c c c }
\hline\hline
\textbf{Generation} & \textbf{Particle} & \textbf{Symbol} & \textbf{Mass} $\left(m_{e}\right)$ & \textbf{Charge} $\left(e\right)$ \\
\hline
1  & Electron & $\textrm{e}$        &  $1.00$    & $-1$  \\
2  & Muon     & $\mu$               &  $207.67$  & $-1$  \\
3  & Tau      & $\tau$              &  $3477.00$ & $-1$  \\
\hline
 & & & \textbf{Energy} ($\textrm{eV}$) & \\
\hline
1  & Electron-neutrino & $\nu_{e}$        &  $2.20\times10^{0}$    & $0$  \\
2  & Muon-neutrino     & $\nu_{\mu}$               &  $1.70\times10^{5}$  & $0$  \\
3  & Tau-neutrino      & $\nu_{\tau}$              &  $1.55\times10^{7}$ & $0$  \\
\hline\hline
\end{tabular}
\end{table}


\begin{figure}
\centering
\epsfysize=6.0cm
\epsfbox{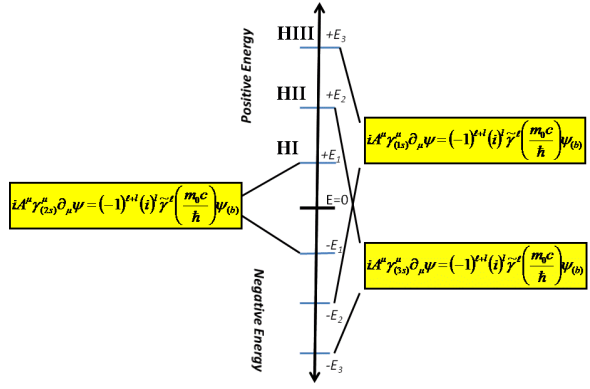}

\caption[Energy Level Diagram from the Curved Spacetime Dirac Equation without the Cosmology 4-Vector Field]{\figtextfont Energy level diagram showing the expected energy levels from the three Dirac Curved Spacetime Equations. This energy level diagram does not include the Cosmological $\textit{4}$-Vector Field proposed in Nyambuya ($2009a$).}
\label{energyls}
\end{figure}

Given that $m_{e}<m_{\mu}<m_{\tau}$ where $m_{e}$, $m_{\mu}$ and $m_{\tau}$ are the mass of the Electron, Muon and the Tau-particle respectively, it follows that if the above is the correct, then the three-member hierarchy of these fermions finds a solution \textit{vis}, why it exists. It is important to say, this argument has been made on the assumption of Dirac's hypothesis that the negative energy states are filled thus not observable. Otherwise if they where observable we would have instead of a three member hierarchy, a six member hierarchy that includes negative energy as shown in the table (\ref{lep-el}). The energy level structure emerging from the there curved spacetime equations is shown in figure (\ref{energyls}). In the reading Nyambuya ($2009a$), we shall add a $\textit{4}$-Vector Cosmological Field to the current curved spacetime equations and this will change the energy level structure as this modification will bring about dark particles. 

Back to the present reading; in  table (\ref{lep-el}), the negative energy Electron has been named Nelectron -- we simple added an N at the beginning of the name of the positive energy counterpart, likewise we shall do so with all other negative energy particles, their names will all start with an N followed by the name of their positive energy counterpart. Unlike Dirac, there is no need for us to worry about negative energies as these won't pause to us the same problem that they did to Dirac -- this will be demonstrated in Nyambuya ($2009a$). So the reader should not worry about these at the moment, we shall proceed smoothly without panic.

\begin{table}[!h]
\centering
\caption{\tabletitlefont Lepton Energies}\label{lep-el}
\vspace*{0.3cm}
\tabletextfont
\begin{tabular}{c l c l}
\hline
\textbf{ Energy Level} & \textbf{Particle} & \textbf{Symbol} & \textbf{Energy Formula}    \\
\hline\hline
+3  & Tau & $\tau$ & $E=+\sqrt{2\mathcal{P}^{2}c^{2}+\mathcal{M}^{2}_{0}c^{4}}+\mathcal{P}c$         \\
+2  & Muon     & $\mu$ & $E=+\sqrt{2\mathcal{P}^{2}c^{2}+\mathcal{M}^{2}_{0}c^{4}}-\mathcal{P}c$                \\
+1 & Electron      & $\textrm{e}$ & $E=+\sqrt{\mathcal{P}^{2}c^{2}+\mathcal{M}^{2}_{0}c^{4}}$              \\
\hline
-1  & Nelectron & $\textrm{e}_{N}$ & $E=-\sqrt{\mathcal{P}^{2}c^{2}+\mathcal{M}^{2}_{0}c^{4}}$         \\
-2  &  Nmuon    &  $\mu_{N}$ & $E=-\sqrt{2\mathcal{P}^{2}c^{2}+\mathcal{M}^{2}_{0}c^{4}}+\mathcal{P}c$                  \\
-3  &   Ntau    & $\tau_{N}$ & $E=-\sqrt{2\mathcal{P}^{2}c^{2}+\mathcal{M}^{2}_{0}c^{4}}-\mathcal{P}c$                 \\
\hline\hline
\end{tabular}
\end{table}

Through we have not pinned down the problem conclusively, from the above, we can safely say with a high degree of certainty that, if the Curved Spacetime Dirac Theory is correct -- as we believe it is; then, the origins of the mass hierarchy in the form of the  three generations, arise because spacetime has three states of curvature, either it is \textit{quadratic curve}, \textit{parabolically curved} or \textit{hyperbolically curved}. Each of these curvature states gives rise to its own Curved Spacetime Equation whose energies solutions are different from for the particles of there other curved spacetime, hence the three generations.

\section{\sectionfont Neutrinos\label{netrino}}  

If we set $\mathcal{M}_{0}=0$, it follows that for all the equations describing the particles, \textit{i.e.}, the QST-Eqns, HST-Eqns and PST-Eqns, these equations reduce to just three equations which, written in the compact notation,  are given by:

\begin{equation}
i\hbar A^{\mu}\gamma^{\mu}_{(ks)}\partial_{\mu}\psi=0,\label{pneutrino}
\end{equation}

and the corresponding energy formula we simple insert $\mathcal{M}=0$ into (\ref{energy2}) and so doing we obtain:

\begin{equation}
E(\nu)=\left(\lambda \pm \sqrt{\left(1+\lambda^{2}\right)}\right)\mathcal{P}c\label{energyn}.
\end{equation}

\textit{If and only if}, the suggestion made in equation (\ref{mq}), that the rest-mass is directly proportional to the electronic charge of a particle, then, these equations may-well describe neutrinos since neutrinos are spin-1/2 particles having a zero electronic charge as are the particles described by equations (\ref{pneutrino}) for the case $s=1$. As with the case of the leptons, these particles will exhibit the same three-level hierarchy which is expect to to follow the order [$E_{e}(\nu)<E_{\mu}(\nu)<E_{\tau}(\nu)$]  where these energies would represent the neutrinos in the same order \textit{i.e.,} $\nu_{\tau}$, $\nu_{\mu}$ and $\nu_{e}$.

\begin{table}[!h]
\centering
\caption{\tabletitlefont Neutrino Energy Levels}\label{netrino}
\vspace*{0.3cm}
\tabletextfont
\begin{tabular}{c l c l}
\hline
\textbf{Energy Level} & \textbf{Particle} & \textbf{Symbol} & \textbf{Energy Formula}    \\
\hline\hline
+3  & Tau-neutrino        & $\nu_{\tau}$ & $E(\nu)=+\left(\sqrt{2}+1\right)\mathcal{P}c$         \\
+2  & Muon-neutrino       & $\nu_{\mu}$ & $E(\nu)=+\left(\sqrt{2}-1\right)\mathcal{P}c$                \\
+1  & Electron-neutrino   & $\nu_{e}$ & $E(\nu)=+\mathcal{P}c$              \\
\hline
-1  & Nelectron-neutrino & $\nu^{N}_{e}$ & $E(\nu)=-\mathcal{P}c$         \\
-2  &  Nmuon-neutrino    & $\nu^{N}_{\mu}$ & $E(\nu)=-\left(\sqrt{2}-1\right)\mathcal{P}c$                  \\
-3  &   Ntau-neutrino    & $\nu^{N}_{\tau}$ & $E(\nu)=-\left(\sqrt{2}+1\right)\mathcal{P}c$                 \\
\hline\hline
\end{tabular}
\end{table}

From the above, one finds that  the ratio of the energy of  $\nu_{\tau}$ and  $\nu_{\mu}$ yield a constant ratio, \textit{i.e.}:

\begin{equation}
\frac{E_{\tau}(\nu)}{E_{\mu}(\nu)} =\frac{\sqrt{2}+1}{\sqrt{2}-1}\simeq 5.83,\label{nu-ratio}
\end{equation}

Checking the observed ratio of the energy of $\nu_{\tau}$ and $\nu_{\mu}$, we find that $E(\nu_{\tau})/E(\nu_{\mu})=91.2$. This is significant disagreement!  Is there a way to modify the theory to circumvent this discrepancy and reconcile it with observations. There seems to be such and avenue to reconcile with observations; we  can make a modification to the theory by the addition if of a constant to the energy term. This modification does not alter the essence of the theory. We shall make this modification by adding a universal constant to the energy equation (\ref{energy2}), \textit{i.e.,} $E\longmapsto E+\Lambda \hbar c$ thus leading to $E=\Lambda \hbar c+\lambda \mathcal{P}c\pm \sqrt{\left(1+\left|\lambda\right|\right)\mathcal{P}^{2}c^{2}+\mathcal{M}^{2}_{0}c^{4}}$. With this, we will have (\ref{nu-ratio}) being given by:

\begin{equation}
\frac{E_{\tau}(\nu)}{E_{\mu}(\nu)} =\frac{\sqrt{2}+1+(\Lambda \hbar c/\mathcal{P}c)}{\sqrt{2}-1+(\Lambda \hbar c/\mathcal{P}c)},\label{nu-ratio1}
\end{equation}

and from this, it is foreseeable that with an appropriate value of $\Lambda \hbar c/\mathcal{P}c$, one can actually match the observations. This modification of adding $\Lambda \hbar c$ will be done in Nyambuya ($2009a$) and as will be seen therein, this modification goes far beyond just trying to make sure that the theory matches up with experience.
 
The question of whether neutrinos have mass (rest-mass) is at present a ``hot'' topic. If they travel at the speed of light, as they seem, then according to the Special Theory o\textit{f} Relativity, they must have a zero-rest-mass. The Standard Model of particle physics assumes that they are massless (zero rest-mass), although adding massive neutrinos to the basic framework is not difficult and this is sometimes what is done. The need for neutrino mass, comes in from the experimentally established phenomenon of neutrino oscillation (a phenomenon where the neutrino switches flavors, \textit{i.e.,} $\nu_{e}\leftrightarrow\nu_{\mu}$, $\nu_{e}\leftrightarrow\nu_{\tau}$ and $\nu_{\tau}\leftrightarrow\nu_{\mu}$) which requires if not demand that neutrinos to have nonzero masses rest-mass (see \textit{e.g.} Karagiorgi \textit{et al.} 2007). Neutrino oscillations where detected in 1998 for the first time from the Super-Kamiokande neutrino detector, and  this pointed to the fact that neutrinos may indeed have mass (Fukuda \textit{et al.} 1998). Since then, the question of whether neutrinos have mass has been a top controversy and has never really been settled satisfactorily.

In the present theory, we have been able to show that neutrinos may actually have a zero rest-mass and this is based on the fact that they have a zero electronic charge. Since we pointed out that the rest-mass of a particle ought to be intimately connected to the electronic charge of the fundamental particle in question by the relationship equation (\ref{mq}), from this it flows that neutrinos ought therefore to have a zero rest-mass. If the present theory is correct, then, it means it should be possible using this theory, and maybe other exogenous ideas, to explain neutrino oscillations for massless neutrinos. For this to happen, certainly a more and better understanding of the present theory is needed. Until it has been found that it is possible using the present theory, to explain neutrino oscillations for massless neutrinos, one can not make any bold statements that the neutrinos are massless. It is only interesting that the theory makes in-roads to the endeavor of finding an answer.

\section{\sectionfont Discussion \& Conclusion\label{dis}}

By no means can this reading be considered to be complete -- that it can stand on its own, that we can make bold conclusions from it, \textbf{\textit{no}}, the truth is that \textit{brick-by-brick} we are building on the theory set out in Nyambuya ($2007, 2008$) and this reading is just a part of that building process. We can safely say, we have worked out the final painting of the theory, but we have to carefully let this picture in small quanta, each time checking if this final picture corresponds with experience and also, checking if of the known mysteries and anomalies (\textit{e.g.} darkmatter, darkenergy \textit{etc}), are we able to explain these or does the final picture naturally shade light on these matters -- the answer is yes, some of the anomalies and mysteries appear to be explained by the theory. As to whether or not we are on the right path, we would like to give the reader our true convictions that, as things stand in the present moment, we are in no doubt of this, that we are on the right path.

First, to the three Curved Spacetime Dirac Equation proposed in Nyambuya ($2008$), $189$ more equations have been generated resulting in a total of $192$ equations. These equations have been classified into twelve different classes as shown in table (\ref{equations}). The properties of these equations have been investigated and formally written down. It has been shown that these equations are:

\begin{enumerate}
\renewcommand{\theenumi}{(\arabic{enumi})}
\item$\,$ Invariant under a Lorentz transformation.  Lorentz invariance, is a symmetry these equations can not afford to go against, they must fulfill this if at all they are to be physically valid equations. Unlike in the case of the bare Dirac Theory, all the $\gamma$-matrices are pure constants, they do not depend on the frame of reference.
\\
\item$\,$ The equations are invariant under charge conjugation and from this, it has been shown that this implies an intimate relationship between rest-mass and electronic charge. This relationship, suggest that the rest-mass is directly proportional to the electronic charge (equation \ref{mq}). If this is correct, then, it could mean neutrinos should have a zero rest-mass.
\\
\item$\,$ The CST-Eqns violate $\textrm{T}$ and $\textrm{P}$-symmetries and as-well $\textrm{CP}$ and $\textrm{CT}$  but obeys the combination $\textrm{CPT}$ invariant. If antiparticles have their hand of time in the opposite direction to the forward as in the Dirac Theory, then one finds an explanation to the mystery of why the Universe is dominated by matter with little if any antimatter in it. We have not used this property of the equations to say this should be the explanation of the mystery of the dominance of matter over antimatter. We are aware of more subtle and robust explanation and this if found in Nyambuya (2007) but there is need to coalesce this with the idea of a $\textit{4}$-Vector Cosmological Field proposed in Nyambuya ($2009a$) before one gives a final solution to this. One can safely say that the generations obseved in Leptons and Quarks, are a result of the three states of curvature of spacetime.
\\
\item$\,$ It has been shown that the Curved Spacetime Dirac Equations naturally exhibit a three level hierarchy in their energy solutions and this is a result of the three states of the curvature of spacetime, \textit{i.e.}, spacetime can either be Quadratically, Hyperbolically or Parabolically curved and each of these states has their own Dirac Equation. These three Dirac Equations, seem to strongly point to an explanation as to why and how we come to have Leptons exhibiting a similar mass hierarchy. Pending the outcome of Nyambuya ($2009a$), we have avoided to conclude that this three level hierarchy explains the \textit{Lepton Generation Problem} because we suspect there may be more leptons that this theory predicts. Surely, this finding is an important finding that needs to pursued to its end.

\end{enumerate}

The above four points are just about the major highlights emerging from the present reading. The original intent when this reading was conceived was to do only cover the points highlighted in $1$ and $3$. The rest of the findings where unexpected. We did not even expect to end-up with $192$ equations! These $192$ equations derived here can -- as already shown in the main body of this reading; be condensed into one compact equation, namely  equation (\ref{aldiracs}). From its very design,  equation (\ref{aldiracs}) is applicable to all particles of spin whose magnitude is equal or greater than $1/2$. Despite these equations being so many, they fall into a family of three equations and all of them can be understood from by understanding the three original equations (\ref{dirac1},\ref{dirac2},\ref{dirac3}) albeit in their modified form -- this greatly simplifies matters.

Second, the intimate relationship between rest-mass and electronic charge that is implied by the present theory suggests that neutrinos may be massless. We have suggested that there must exist a direct proportionality relationship between rest-mass and electronic charge and if this is the case, then any fundamental particle of non-zero electronic charge will have a finite rest-mass, hence given that neutrinos have no net electronic charge, they will accordingly, have a zero-rest-mass! If an answer can be found, \textit{i.e.}, a definite answer to this problem of whether neutrinos do indeed have mass, it is a milestone for physics. The current Standard Model of particle physics assumes that neutrinos are massless, although sometimes this assumption is dropped. 

Third, accepting that neutrinos are massless requires us to modify the theory to include an all pervading and permeating cosmic energy-momentum field and this task is underway in the on-going reading, Nyambuya ($2009a$). This energy-momentum, may serve as the typically assumed Cosmological Constant and its magnitude appears at present like it can not be calculated from the present theory. If this modification brings the neutrino model from the present theory to match the observations, this would point to the correctness of the theory the meaning of which would strongly suggest that neutrinos should be mass-less, and if neutrinos are massless, then, it should be possible using the present theory to explain neutrino oscillations. How one would go about this, eludes me at present. One can only hope that as a better understanding dawns, light on this matter will dawn too.

In-closing, allow me to say that, we do not know to which journal we are going to send the final completed reading of Nyambuya ($2009a$). This reading should finish the present. But one thing is clear, the preprints will be available on my \textit{viXra.org} profile and also on my \textit{arXiv.or}g profile.

\end{document}